\documentclass[structabstract]{aa}  
\usepackage{natbib}
\usepackage{graphicx}
\usepackage{subfigure}
\usepackage{epsfig}
\usepackage{txfonts}

\def\mgii{Mg{\,\texttt{\scriptsize II}} }

\def\ciii{C{\,\texttt{\scriptsize III]}} }

\def\civ{C{\,\texttt{\scriptsize IV}} }

\def\hbeta{H$\beta$}

\begin{document}
\bibliographystyle{apalike}
   \title{Quasi--stellar objects in the ALHAMBRA survey
	  \thanks{Based on observations collected at the German-Spanish
            Astronomical center, Calar Alto (Almeria, Spain), jointly operated
            by the Max-Planck-Institut f\"ur Astronomie at Heidelberg and the
            Instituto de Astrof\'isica de Andaluc\'ia (CSIC).}}

   \subtitle{I. Photometric redshift accuracy based on 23 optical-NIR filter photometry}

   \author{I. Matute \inst{1}
          \and
          I. M\'arquez \inst{1}
	  \and	
	  J. Masegosa \inst{1} 
	  \and
	  C., Husillos \inst{1}
	  \and
	  A., del Olmo \inst{1}
	  \and
	  J., Perea \inst{1}
	  \and
	  E.\,J. Alfaro \inst{1}
	  \and
	  A. Fern\'andez-Soto  \inst{2}
	  \and
	  M. Moles \inst{1,3}
	  \and
	  J.\,A.\,L. Aguerri  \inst{4}
	  \and  
	  T. Aparicio--Villegas \inst{1} 
	  \and
	  N. Ben\'itez \inst{1}
	  \and
	  T. Broadhurst  \inst{5}
	  \and
	  J. Cabrera--Cano  \inst{1,6}
	  \and
	  F.\,J. Castander \inst{7}
	  \and
	  J. Cepa  \inst{4,8}
	  \and
	  M. Cervi\~no  \inst{1}
	  \and
	  D. Crist\'obal-Hornillos \inst{1,3}
	  \and
	  L. Infante \inst{9}
	  \and  
	  R.\,M. Gonz\'alez Delgado  \inst{1}
	  \and
	  V. J. Mart\'inez \inst{10,11}
	  \and
	  A. Molino \inst{1}
	  \and
	  F. Prada \inst{1}
	  \and
	  J. M. Quintana  \inst{1}
          }

   \institute{Instituto de Astrof\'isica de Andaluc\'ia (CSIC),
              Glorieta de la Astronom\'ia s/n, E--18008 Granada, Spain\\
              \email{matute,\,isabel,\,pepa,\,cesar,\,chony,\,jaime,\,emilio,\,benitez,\,mcs,\,rosa,\,amb,\,fprada,\,quintana@iaa.es}
	 \and
	    Instituto de F\'isica de Cantabria (CSIC-UC), E--39005, Santander, Spain; fsoto@ifca.unican.es 
         \and
	    Centro de Estudios de F\'isica del Cosmos de Arag\'on  (CEFCA), E--44001 Teruel, Spain; moles, dch@cefca.es
	  \and	
	    Instituto de Astrof\'isica de Canarias, La Laguna, Tenerife, Spain; jalfonso@iac.es
	 \and
	    School of Physics and Astronomy, Tel Aviv University, Israel; tjb@wise.tau.ac.il
	 \and
	    Facultad de F\'isica. Departamento de F\'isica At\'omica, Molecular y Nuclear. Universidad de Sevilla, Sevilla, Spain; jcc-famn@us.es
	 \and
	    Institut de Ci\`encies de l'Espai, IEEC-CSIC, Barcelona, Spain; fjc@ieec.fcr.es
	 \and	
	    Departamento de Astrof\'isica, Facultad de F\'isica, Universidad de la Laguna, Spain; jcn@iac.es
	 \and
	    Departamento de Astronom\'ia, Pontificia Universidad  Cat\'olica, Santiago, Chile; linfante@astro.puc.cl
	  \and
	    Departament d'Astronom\'ia i Astrof\'isica, Universitat de Val\`encia, Valencia, Spain; vicent.martinez@uv.es
	  \and 
	    Observatori Astron\`omic de la Universitat de Val\`encia, Valencia, Spain
             }

   \date{Received ?? ??, 2011; accepted ?? ??, 2011}


  \abstract
   {Even the spectroscopic capabilities of today's ground and space--based
     observatories can not keep up with the enormous flow of detections
     ($>10^5$ deg$^{-2}$) unveiled in modern cosmological surveys as: $i$)
     would be required enormous telescope time to perform the spectroscopic follow-ups and
     $ii$) spectra remain unattainable for the fainter detected population.  In
     the past decade, the typical accuracy of photometric redshift (photo-$z$) determination has
     drastically improved. Nowdays, it has become a perfect complement to
     spectroscopy, closing the gap between photometric surveys and their
     spectroscopic follow-ups. The photo-$z$ precision for active galactic nuclei (AGN) has always lagged
     behind that for the galaxy population owing to the lack of proper templates and
     their intrinsic variability.} 
   {Our goal is to characterize the ability of the Advanced Large,
    Homogeneous Area Medium-Band Redshift Astronomical (ALHAMBRA) survey in assigning
     accurate photo-$z$'s to broad-line AGN (BLAGN) and quasi-stellar objects (QSOs) based on their
     ALHAMBRA \textit{very--low--resolution} optical--near-infrared spectroscopy. This
     will serve as a benchmark for any future compilation of ALHAMBRA selected QSOs
     and the basis for the statistical analysis required to derive
     luminosity functions up to $z\sim5$.}
    {We selected a sample of spectroscopically identified BLAGN and QSOs
      and used a library of templates (including the SEDs of AGN, and
      both normal and starburst galaxies, as well as stars) to fit the 23 photometric data
      points provided by ALHAMBRA in the optical and near-infrared (20 medium--band
      optical filters plus the standard \textit{JHKs}).}
   {We find that the ALHAMBRA photometry is able to provide an accurate
     photo-$z$ and spectral classification for $\sim$88\% of the 170
     spectroscopically identified BLAGN/QSOs over 2.5\,deg$^2$ in different
     areas of the survey and brighter than $m_{678}=23.5$ (equivalent to
     $r_{\mathrm{SLOAN}}\sim24.0$). The derived photo-$z$ accuracy is below
     1\% and is comparable to the most recent results in other cosmological
     fields that use photometric information over a wider wavelength
     range. The fraction of outliers ($\sim$12\%) is mainly caused by the
     larger photometric errors for the faintest sources and the intrinsic
     variability of the BLAGN/QSO population. A small fraction of outliers may
     have an incorrectly assigned spectroscopic redshift.
    }
   {The definition of the ALHAMBRA survey in terms of the number of filters,
     filter properties, areal coverage, and depth is able to provide photometric
     redshifts for BLAGN/QSOs with a precision similar to any previous survey
     that makes use of medium-band optical photometry. In agreement with
     previous literature results, \textnormal{our analysis also reveals that,
       in the $0<z<4$ redshift interval, very accurate photo-$z$ can be
       obtained without the use of near-infrared (NIR) broadband photometry at the expense
       of a slight increase in the outliers. The importance of NIR data is expected to
       increase at higher $z$ ($z>4$). These results are relevant for the
       design of future optical follow-ups of surveys containing a large fraction of
       BLAGN, such as many X--ray or radio surveys.}
}

   \keywords{cosmology: Observations -- galaxies: active -- galaxies:
     distances and redshifts galaxies: evolution -- galaxies: high-redshift -- 
     quasars: general
               }

   \maketitle
%

\section{Introduction}

The role of active galactic nuclei (hereafter AGN) in the formation of the
early structures and their later evolution has been reviewed over the past 15
years, becoming a key ingredient of galaxy evolution models 
(e.g. Cattaneo 2002; Menci et al. 2003; Croton et al. 2006; Hopkins et al. 2010 and references therein). 
Evidence shows that many, if not all, massive galaxies harbor
supermassive black holes (SMBHs; e.g. Kormendy \& Richstone 1995). The close
interaction between the formation and growth of the SMBH and the evolution of
its host galaxy were initially revealed by: $i$) the tight correlations
between the masses of central SMBHs and the velocity dispersions and
luminosities of the bulges of many galaxies (Tremaine et al. 2002); $ii$) the
remarkable similarities in the redshift at which starburst and accretion
activities ocurred and $iii$) the observation of the so--called \textit{downsizing}
effect not only for the galaxy population but also for AGN, i.e. the most massive
galaxies appear to have assembled the majority of their stars earlier than lower mass
galaxies (Cowie et al. 1996; Zheng et al. 2007), while the density of low--luminosity 
AGN peaks at lower $z$ than the more powerful ones (e.g. Hasinger
et al. 2008 and references therein).  Therefore, the measure of the space
density of AGN with cosmic time not only provides information about the relative
importance of accretion activity to the global energy output in the universe
but also places important constraints on early structure formation and
galaxy evolution (e.g. Di Matteo et al. 2005, Hopkins et al. 2010).

Quasi-stellar objects (QSOs) are the members of the AGN family that have
particularly high intrinsic luminosities allowing them to be detected at
large distances and to provide unique inside into the early history of the AGN-host
galaxy interaction. Moreover, QSOs are potential contributors to the ultraviolet (UV)
ionizing background (Cowie et al. 2009) and have probably played a
non-negligible role in the reionization of the universe (Fan et al. 2006, Wang
et al. 2010).

The optical selection of QSOs has been performed mainly with follow-up
spectroscopic observations of color--color selected candidates (e.g. SDSS,
Richards et al. 2002 and 2dF, Croom et al. 2004). These observations use
slitless or prism spectroscopic surveys and poorly efficient flux--limited
spectroscopic surveys (e.g. VIMOS--VLT Deep Survey, Gavignaud et al. 2006;
Bongiorno et al. 2007).  A novel technique with respect to previous selection
criteria was introduced by the CADIS (Meisenheimer et al.  1998) and COMBO--17
surveys (Wolf et al. 2003). These photometric surveys used several optical
broad-- and medium--band filters to characterize the nature of the
detected population and derive its photometric redshift (photo-$z$) via the 
spectral energy distributions (SEDs). The fluxes reached by the survey 
have allowed the study of the high-$z$ QSO population, thus overcoming the 
problem of QSO incompleteness in the redshift interval $2.2\le z \le 3.6$ (Richards 
et al. 2002). This redshift range is important because it corresponds to the peak 
and the turnover of the observed QSO space density (e.g. Wolf et al. 2004).

Over the past decade, a clearer understanding of the QSO evolution has been
achieved thanks to a more accurate characterization of their different SEDs, to a
more precise treatment of their variability, and to a significant improvement in their
photo-$z$ determination. This advance has been
encouraged by the conception of modern cosmological surveys and newly
available space-based observing facilities (e.g. \textit{HST},
\textit{XMM--Newton}, \textit{Chandra}, \textit{Spitzer} and \textit{Herschel}
among others) that have been able to detect a large amount of sources
($\ge 10^6$).  In particular, when a given scientific goal does not require
detailed knowledge of the spectral properties of individual objects, properly designed
photometric surveys can provide a highly reliable photo-$z$ and spectral
classification for each source. These photo-$z$'s are an essential complement
to the usually small fraction of sources with spectroscopic redshifts in major
extragalactic surveys and to more reliably probe the fainter detected population,
which is difficult to access using current ground-based spectroscopic
observatories. In addition, photo-$z$ are used to
validate uncertain spectroscopic redshifts typically obtained for spectra of low
signal-to-noise ratio (S/N) or limited wavelength coverage 
(e.g. Fern\'andez-Soto et al. 2001). Several computational methods have been 
developed to derive photometric redshifts with increasingly high precision 
(\textit{BPZ, HyperZ, LePhare, ZEBRA, AnnZ, EAzY}, among others).  
Only recently have photo-$z$ for AGN
(Salvato et al. 2009; Luo et al. 2010; Cardamone et al. 2010) reached
accuracies similar to those computed for normal and starburst galaxies
($\sim$1--2\%; e.g. Ilbert et al 2009).  A description of the current state of
the art photo-$z$ computation as well as a detailed performance comparison of
various photo-$z$ codes was provided by Hildebrandt et al. (2010).

In this context, we present the analysis of photometric redshift solutions
found for a population of spectroscopically identified QSOs using the optical
and near-infrared (NIR) multi-band catalog of the ALHAMBRA survey. The ALHAMBRA survey was
designed with an optimal filter combination in order to provide one of the
most homogeneous, large, deep, and accurate photometric surveys. Given the
proposed depth of the ALHAMBRA filters of AB$\sim$24.5--25, we expect to
sample the whole QSO LF up to $z\sim4.2$ and up to z$\sim$6 for sources with
$M_B>-24.2$. At the current stage, the survey has mapped $\sim$2.5\,deg$^2$ of
the sky in seven different regions. The results presented here, and the comparison
with existing data from other cosmological surveys, will prove the
capabilities of the survey to derive accurate photometric redshifts for the
BLAGN/QSO population. Furthermore, this test will potentially identify
redshift ranges for which QSO photo-$z$ estimation maybe unreliable or
QSOs with atypical SEDs that would then be suitable for more detailed study.

This paper is structured as follows. In Sect.\,2, we describe the current
photometric catalog from the ALHAMBRA survey as well as the ancillary data
available in each of the ALHAMBRA fields from other cosmological surveys. This
section also introduces the QSO sample selection. The methodology followed
during the photo-$z$ determination is discussed in Sect.\,3 while in Sect.\,4 we
quantify the precision of our photo-$z$ estimates by comparing them with previous
results for this type of sources. Finally, Sect.\,5 discusses the implications of
our results and planned future analysis. A detailed QSO catalog will be
presented in a forthcoming paper. Throughout our analysis, we assume a
$\Lambda$CDM cosmology with $H_o=70\,\mathrm{km\,s^{-1}}\,\mathrm{Mpc}^{-3}$,
$\Omega_{\Lambda}=0.7$, and $\Omega_{\mathrm{M}}=0.3$. Unless otherwise
specified, all magnitudes are given in the AB system.


\section{Data set}

%
\begin{table*}[ht]
\caption{The ALHAMBRA fields}             
\label{table:fields}      
\centering                          
\begin{tabular}{c c c c c c c c c }        
\hline\hline                 
Field & alpha(J2000) & delta(J2000) & Area (deg$^2$) & Obs. period & Surveys &
E($B-V$) & Spectro-QSO & Source \\ 
($i$) & ($ii$) & ($iii$) & ($iv$) & ($v$) & ($vi$) & ($vii$) & ($viii$) & ($ix$)\\
\hline                        

  ALH-2 & 02 28 32.0 & +00 47 00  & 0.50 & Sep05--Nov09 & DEEP2          & 0.030  & 30/29 &  1,2\\
  ALH-3 & 09 16 20.0 & +46 02 20  & 0.25 & Dec04--May09 & SDSS           & 0.015  &  2/2  &  1\\
  ALH-4 & 10 00 28.6 & +02 12 21  & 0.25 & Dec04--May09 & COSMOS         & 0.018  & 81/77 &  3\\
  ALH-5 & 12 35 00.0 & +61 57 00  & 0.25 & May05--Jun09 & HDF-N          & 0.011  & 18/15 &  1,4,5\\
  ALH-6 & 14 16 38.0 & +52 25 05  & 0.19 & Aug04--Aug09 & EGS-AEGIS      & 0.011  & 35/33 &  1,2,6\\
  ALH-7 & 16 12 10.0 & +54 30 00  & 0.50 & Aug04--Jul09 & SWIRE/ELAIS-N1 & 0.007  & 11/11 &  1,4,7\\
  ALH-8 & 23 45 50.0 & +15 34 50  & 0.50 & Aug04--Aug09 & SDSS           & 0.024  &  3/3  &  1\\
\hline
  TOTAL QSOs\tablefoottext{a}  &  &     & 2.44 &    	       &   & 180/170\\ 

\hline                                   
\end{tabular}
\tablefoot{\scriptsize \\
  ($i$)= ALHAMBRA field name; ($ii$,$iii$)=Central coordinates of the field; ($iv$)= Area
  covered by each field; ($v$)= Period between the beginning and the end of the
  observations in a given field; ($vi$)= Name of the cosmological survey for which a
  particular ALHAMBRA field overlaps; ($vii$)= Mean Galactic reddening along the line of
  sight derived by Schlegel, Finkbeiner \& Davis (1998) from the $IRAS$ 100\,$\mu$m
  data; ($viii$)= Total number of spectroscopically identified QSOs in the field and those
  within our selection criteria; ($ix$)= Source of the QSO classification and
  spectroscopic redshift: (1)= Schneider et al. (2010) ; (2)= DEEP webpage, Davis et al.
  (2003); (3)=Brusa et al. (2010); (4)= Veron--Cetty, M.\,P. \& Veron, P. (2010); (5)=
  Barger et al. (2008); (6)= C. Willmer \textit{priv. comm}; (7)= Rowan-Robinson et al.
  (2008); \\
\tablefoottext{a}{Sum over all the ALHAMBRA fields of the total/selected spectroscopic
		  QSOs.}
}
\end{table*}
%

\subsection{Photometric data: The ALHAMBRA survey.}

The ALHAMBRA\footnote{http://alhambra.iaa.es:8080} (Advanced, Large,
Homogeneous Area, Medium-Band Redshift Astronomical) survey provides a
photometric dataset over 20 contiguous, equal-width, non-overlapping,
medium-band optical filters (3500 -- 9700\,\AA) plus 3 standard broad-band
NIR filters $J$, $H$, and \textit{Ks} over 8 different regions
of the northern hemisphere (Moles et al. 2008). The survey aims to understand
the evolution of the structures and the different families of extragalactic
sources throughout cosmic time by sampling a large enough cosmological
fraction of the universe. This requires precise photometric redshifts for
several hundreds of thousands objects and, therefore, a survey with high
photometric accuracy as well as deep and wide spectral coverage over a large
area. The simulations of Ben\'itez et al.  (2009), relating the image depth and
photo-$z$ accuracy to the number of filters, have demonstrated that the
filter--set chosen for ALHAMBRA can achieve a photo-$z$ precision, for normal
and star-forming galaxies, that is three times better than a classical 4--5 optical
broad-band filter set. The final survey parameters and scientific goals, as
well as the technical properties of the filter set were described by
Moles et al. (2008). The survey has collected its data for the 20+3
optical-NIR filters in the 3.5m telescope, at the Calar Alto
observatory\footnote{http://www.caha.es}, using the wide--field camera LAICA
in the optical and the OMEGA--2000 camera in the NIR. The full
characterization, description, and performance of the ALHAMBRA optical
photometric system was presented by Aparicio--Villegas et al. (2010). The
strategy of ALHAMBRA for each run has been to observe the fields with the
lowest airmasses trying to complete the requested integration time for each
filter in order to reach the planned depth (ABmag$\sim$25). In consequence,
and depending also on the telescope/instrument downtime and weather, the time
to complete each filter in each field can vary from months to several
years\footnote{Further details of the observations will be provided in a
  companion paper.}.  Therefore, although ALHAMBRA photometry allows us to detect
variability, the lack of a common band(s) taken during all the observing runs
does not allow us to correct for its effect.  The deep NIR counts in one of
the ALHAMBRA fields (ALH-8), over $\sim$0.5 deg$^2$, which has a 50\% detection
efficiency depth of $J \sim22.4$, $H \sim21.3$, and $K_s \sim20.0$ (Vega),
have been analyzed by Crist\'obal--Hornillos et al. (2009). Their results
helped to constrain different type-dependent galaxy evolutionary
models.

\begin{figure}[t]
   \centering
    \includegraphics[width=9.3cm,angle=0, trim=14 0 0 0]{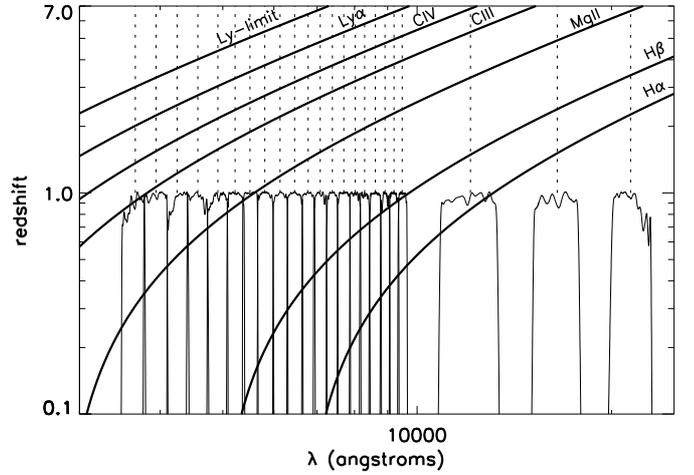}
      \caption{\scriptsize Wavelength coverage of the ALHAMBRA filter set for the
      LAICA camera CCD1 (thin continuous lines). The position of the most prominent QSO
      and BLAGN emission lines, plus the Lyman limit, are  shown evolving with redshift as
      thick lines. All filters have been normalized to unity (see Aparicio-Villegas et
      al. 2010 for their true efficiency). Dotted lines trace each filter central
      wavelength (Table\,\ref{table:ALHfilters}).}
     \label{ALHfiltesr_QSOlines_vs_z}
   \end{figure}

In this work, we used the seven ALHAMBRA fields for which data have been currently 
observed and reduced. The central coordinates of these 7 fields, the area covered by each,
the observing epoch, their coincidence with other cosmological surveys, and
their mean Galactic extinction are detailed in Table \ref{table:fields}. The
total number of spectroscopically identified QSOs, as well as the fraction
selected for our analysis, are given in Col. 7 of Table \ref{table:fields}.
Figure\,\ref{ALHfiltesr_QSOlines_vs_z} and Table \ref{table:ALHfilters} detail
the general characteristics of the ALHAMBRA 23 filter set.

The photometric data points used in this work are given by the
\texttt{MAG\_AUTO} measure by SExtractor (Bertin \& Arnouts 1996). To
avoid the excessive weight of some points in the computation of a photo-$z$, we
adopted a minimum photometric error of $\delta m = 0.05$ (i.e. if a
photometric error is smaller than 0.05, it is set to 0.05) for the method
described in Sect.\,3. In agreement with the results of
other authors (e.g. Bolzonella et al. 2000), we found that there is no gain,
or even that we obtain poorer results for some objects, when we consider $\delta m <
0.05$. The photometric data points of each object were corrected for
interstellar extinction using the values of $E(B-V)$ provided by the maps of Schlegel,
Finkbeiner \& Davis (1998), which are based on \textit{IRAS} 100\,$\mu$m
data\footnote{http://irsa.ipac.caltech.edu/applications/DUST/}.

\subsection{Sample selection}

Broad-line AGNs (BLAGN hereafter) and QSOs are powerful emitters over the
entire electromagnetic spectrum. They show significant spectral features in
the form of intense emission lines (with EW ranging from several tens to
several thousands of \AA) in the rest-frame UV, optical, and NIR regime. These
properties make QSOs easily detectable out to very high redshifts ($z\sim6$)
and perfect candidates to help improve our understanding of the accretion mechanisms
within SMBHs ($M_{BH}>10^6\,M_{\odot}$). They also probe the distribution 
of large-scale structures and the
physical conditions of the intergalactic medium (IGM). The strong features
that characterize the QSO optical emission spectrum allow us to test the ALHAMBRA
photometry and its ability to produce \textit{very low resolution
spectra}. This would provide a correct spectral classification and a high-precision 
redshift estimate for the expected population of several thousands of QSOs.

We selected our initial QSO candidates from a subsample of the current
ALHAMBRA catalog (v3), which was created using the following photometric criteria:

\begin{itemize}
  \item A survey quality flag $\ge0.7$. Each ALHAMBRA source was flagged
    with a parameter (``percent--weight'' in the catalog) that takes into
    account the total exposure time of a given source relative to the maximum
    for a given field. A low value of this flag ($<0.70$) indicates that the
    source is either within a region strongly affected by the dithering process
    during the observation, contains bad pixels/artifacts, or is located
    near a bright (masked) source. A detailed comparison with the deeper
    photometry data available for some fields from other surveys shows that source
    detections with percent--weight$\ge0.70$ are highly reliable and that a negligible
    fraction of them are spurious.
  \item The source must be within the very high confidence magnitude interval
    of the survey.  The chosen magnitude of reference is A678M filter centered
    on 6789\,$\AA$ and the interval is defined by $17.0\le A678M \le23.5$. The
    bright magnitude cut ensures that no source saturates any filter, while the
    faint cut avoids sources with photometric errors larger than $\simeq$0.2
    magnitudes.

\end{itemize}

We decided against the inclusion of a stellarity criteria, as the precision of
the one derived by the SExtractor (Bertin \& Arnouts 1996) package was valid
only for the brighter part of the QSO sample ($m_{678} \le 22$).

%
\begin{table}
\caption{ALHAMBRA filter characteristics}             
\label{table:ALHfilters}      
\centering                          
\begin{tabular}{c c c c c c c}        
\hline\hline                 
Name  &   $\lambda_{mean}$ & FWHM  & AB  & Offset & 
$\langle m \rangle$ &$\langle \sigma(m)\rangle$ \\    
       &  ($\mu$m) & ($\mu$m)& corr. &  & (AB) &  \\
 (1) & (2) & (3) & (4) & (5) & (6) & (7) \\
\hline 
A366M & 0.3661 & 0.0279 &  0.96 & -0.033 & 21.81 & 0.08 \\ 
A394M & 0.3941 & 0.0330 &  0.02 & -0.210 & 21.60 & 0.05 \\ 
A425M & 0.4249 & 0.0342 & -0.13 & -0.081 & 21.65 & 0.05 \\ 
A457M & 0.4575 & 0.0332 & -0.18 & -0.011 & 21.62 & 0.07 \\ 
A491M & 0.4913 & 0.0356 & -0.05 & -0.065 & 21.54 & 0.06 \\ 
A522M & 0.5224 & 0.0326 & -0.04 & -0.054 & 21.48 & 0.06 \\ 
A551M & 0.5510 & 0.0297 &  0.01 &  0.003 & 21.48 & 0.07 \\ 
A581M & 0.5809 & 0.0324 &  0.07 & -0.001 & 21.39 & 0.05 \\ 
A613M & 0.6134 & 0.0320 &  0.13 &  0.009 & 21.35 & 0.06 \\ 
A646M & 0.6461 & 0.0357 &  0.23 &  0.006 & 21.33 & 0.08 \\ 
A678M & 0.6781 & 0.0314 &  0.24 & -0.046 & 21.20 & 0.06 \\ 
A708M & 0.7078 & 0.0332 &  0.29 & -0.055 & 21.14 & 0.05 \\ 
A739M & 0.7392 & 0.0304 &  0.34 &  0.007 & 21.16 & 0.06 \\ 
A770M & 0.7699 & 0.0354 &  0.39 &  0.000 & 21.11 & 0.06 \\ 
A802M & 0.8020 & 0.0312 &  0.44 &  0.002 & 21.06 & 0.07 \\ 
A829M & 0.8294 & 0.0296 &  0.48 &  0.007 & 21.00 & 0.08 \\ 
A861M & 0.8614 & 0.0369 &  0.54 & -0.023 & 20.91 & 0.05 \\ 
A892M & 0.8918 & 0.0303 &  0.50 &  0.022 & 20.93 & 0.08 \\ 
A921M & 0.9208 & 0.0308 &  0.48 &  0.028 & 20.83 & 0.10 \\ 
A948M & 0.9482 & 0.0319 &  0.52 &  0.077 & 20.65 & 0.15 \\
$J$   & 1.2094 & 0.2471 &  0.87 &  0.104 & 20.65 & 0.06 \\ 
$H$   & 1.6482 & 0.2665 &  1.38 &  0.186 & 20.41 & 0.08 \\ 
$Ks$  & 2.1409 & 0.3040 &  1.83 &  0.155 & 20.20 & 0.09 \\                                
\hline                        
\hline                             
\end{tabular}
\tablefoot{\scriptsize \\
Columns: (1) Filter name; (2) Filter mean wavelength; (3) Filter FWHM; (4)
AB--Vega magnitude correction: $m_{\mathrm{AB}}$ = $m_{\mathrm{Vega}} +
$AB\_correction; (5) Offsets applied to each filter as
$m_{\mathrm{final}}$=$m_{\mathrm{filter}} + $offset during the photometric
redshift determination (see Sect.\,3.3 for details); (6) Mean magnitude in each
filter band for the spectroscopic sample (Sect.\,\ref{spectro-data}); (7)
Mean magnitude errors in each filter band for the spectroscopic sample. \\
}
\end{table}
%

\subsection{Spectroscopic data}
\label{spectro-data}

To assess the quality and accuracy of the photo-$z$
determination for the ALHAMBRA database, we compiled all the published or
publicly available spectroscopic information for BLAGNs/QSOs. The online
services and public spectroscopic catalogs included the Sloan Digital Sky
Survey (SDSS\footnote{http://www.sdss.org/}) DR7 (Schneider et al. 2010), the
Deep Extragalactic Evolutionary Probe
(DEEP/DEEP2\footnote{http://deep.ucolick.org}; Davis et al. 2003), the
All-wavelength Extended Groth strip International Survey
(AEGIS\footnote{http://aegis.ucolick.org/}; Davis et al. 2007), the
COSMOS\footnote{http://cosmos.astro.caltech.edu/} XMM source catalog (Brusa
et al. 2010), the GOODS-North redshift compilation by Barger et al. (2008),
the SWIRE\footnote{http://swire.ipac.caltech.edu/swire/swire.html}
spectroscopic catalog by Rowan-Robinson et al. (2008) and the 13th edition of
the Veron-Cetty \& Veron QSOs catalog (2010;
VERONCAT\footnote{http://heasarc.gsfc.nasa.gov/W3Browse/all/veroncat.html}
hereafter).  We verified the quality of the identifications according to the
following criteria: $i$) All spectra from the DR7 SDSS catalog were visually
inspected to determine whether they contained broad-line emission; $ii$) spectra with a high-quality
classification flag (flag$\ge$3) were selected from the DEEP/DEEP2 and the
AEGIS database and visually inspected to confirm that they displayed broad-line emission; $iii$) as
neither a spectral classification nor a redshift quality were given by Barger et al. in
the GOODS-N field, valid BLAGN/QSO candidates were selected based on their
hard X--ray luminosity ($L_X$[2--8\,kev]) being brighter that
$10^{43}$\,erg\,s$^{-1}$; $iv$) in the COSMOS field, we selected the high-quality 
public spectra of BLAGN (flags\footnote{flags 18 and 218 refer to a
  spectroscopic redshift computed with a single line} 13, 14, 18, 213, 214, 218; Lilly et
al. 2007), while we considered as \textit{bona fide} BLAGN/QSO the remaining of XMM
sources without public data reported by Brusa et al. (2010) as 'bl' based on MMT
and IMACS spectroscopy; $v$) all SWIRE sources have high quality spectra
(I. P\'erez--Fourn\'on \textit{priv. comm.}; and $vi$) all the sources from
the VERONCAT were considered as \textit{bone fide} BLAGN/QSOs. No lower redshift
or absolute magnitude cutoff was included in the source selection as our goal
is to test the efficiency of our method and photometry as good redshift
estimators not only for the most powerful BLAGNs and QSOs but also for the low
redshift Seyfert 1 nuclei, which may provide an important contribution to the total light of their
host galaxies.  In all cases, the match between the ALHAMBRA photometry and the
spectroscopic catalogs was performed using a one arcsec search radius and always
identified a unique counterpart.  The sources of the BLAGN/QSO spectroscopic
redshifts for each of the ALHAMBRA fields are:

   \begin{figure}[t]
   \centering
    \includegraphics[width=6.6cm,angle=90]{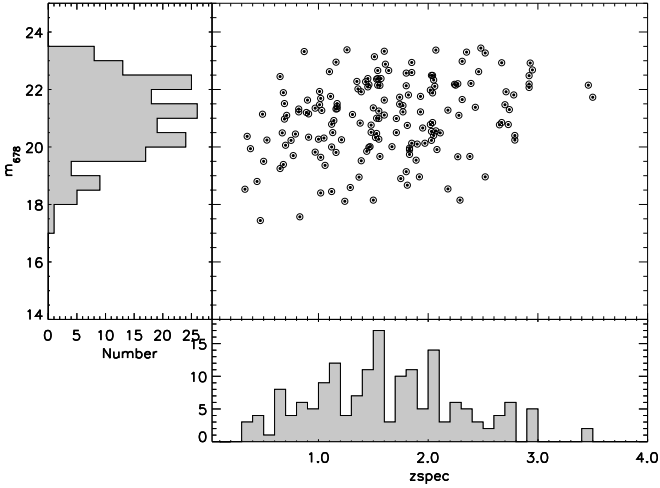}
      \caption{\scriptsize Magnitude--redshift distribution of the 
        selected spectroscopic sample.}
     \label{z_distrib}
   \end{figure}

\begin{itemize}

 \item ALH-2: This $\sim$0.5 deg$^2$ field partially overlaps the deep
   strip of SDSS whose DR7 version provides redshifts for 23 QSOs. The common
   region of this area with field--4 of the DEEP/DEEP2 survey yields 7
   additional sources from their data product release 3
   (DR3\footnote{http://deep.berkeley.edu/DR3/dr3.primer.html}) of the DEEP2
   spectroscopic catalog. Of the 30 spectroscopic redshifts for BLAGN/QSO
   available in the field, 29 (6 from DEEP2 and 23 from SDSS) comply with our
   photometric criteria.

 \item ALH-3: This consists of a $\sim$0.25\,deg$^2$ field area which partially overlaps
   with that of the SDSS. The matching between ALHAMBRA and SDSS DR7 spectroscopy yields 2 QSO
   redshifts, for which both sources verify our photometric criteria.

 \item ALH-4: This is a $\sim$0.25\,deg$^2$ field included in the COSMOS survey
   area. The common area contains a total of 81 BLAGN/QSO redshifts of 
   which 77 (7 from SDSS and 70 from COSMOS) comply with our photometric criteria.

 \item ALH-5: This field covers a $\sim$0.25\,deg$^2$ area overlapping that of
   the GOODS--N.  There are a total of 18 BLAGN/QSOs with spectroscopic
   redshifts (9 from SDSS, 6 from Barger et al. 2008, and 3 from the
   VERONCAT). Of these, 15 (7 from SDSS, 6 from GOODS--N, and 2 from the
   VERONCAT) comply with our photometric criteria.

 \item ALH-6: This $\sim$0.19\,deg$^2$ field is centered on the GROTH strip
   and therefore overlaps with the DEEP2, AEGIS, and SDSS surveys. In total,
   there are 35 spectroscopically identified BLAGN/QSO in this field, of which
   33 sources (6 SDSS, 6 DEEP2 and 21 AEGIS) comply with our photometric criteria.

 \item ALH-7: A $\sim0.5$\,deg$^{-2}$ field centered on the ELAIS-N1 of the
   SWIRE survey. Sources with spectroscopic redshifts and classifications are
   provided by the catalog of Rowan-Robinson et al. (2008), the SDSS
   spectroscopy and the VERONCAT. All 11 BLAGN/QSO found in this field (1 from
   SDSS, 8 from SWIRE, and 2 from VERONCAT) comply with our photometric criteria.

 \item ALHAMBRA-8: The 3 QSO spectroscopic identifications in this
   $\sim0.5$\,deg$^2$ field are provided exclusively by the QSO DR7 catalog of
   the SDSS. All 3 sources comply with our photometric criteria.

\end{itemize}

The final spectroscopic catalog of the ALHAMBRA fields includes 94\% (170/180)
of the total numbers of sources spectroscopically identified in the
different fields. The ALH4-COSMOS field contains $\sim$44\% of the
sources, followed by the ALH2 and ALH6 fields (DEEP2/AEGIS) which corresponds to $\sim$17\%
and 19\% of the IDs, respectively. Table \ref{table:fields} details the number
of identified sources in each field, while Figure\,\ref{z_distrib} shows the
redshift distribution of the selected spectroscopic sample.

Although the surveys from which the spectroscopic sample is extracted encompass a wide
range of selection criteria (color selection in the SDSS, optical flux-limited
in zCOSMOS\_bright, X--ray selected for IMACS and MMT spectroscopy, etc), 
we find their redshift distribution compatible within the errors. The
mean redshift and 1$\sigma$ dispersion for the sources extracted from the
different spectroscopic catalogs are: 1.61$\pm$0.63 (SDSS), 1.99$\pm$0.68
(zCOSMOS\_faint), 1.92$\pm$0.64 (zCOSMOS\_bright), 1.24$\pm$0.45 (MMT),
1.55$\pm$0.66 (IMACS), 1.81$\pm$0.81 (GOODS-N), 1.56$\pm$0.88 (DEEP2),
1.56$\pm$0.63 (AEGIS), 1.42$\pm$0.69 (SWIRE), and 1.24$\pm$0.78 (VeronCat).

%
\begin{table}
\caption{Extragalactic template library.}             
\label{table:templates}      
\centering                          
\tiny
\begin{tabular}{c c c }        
\hline\hline                 
Index & SED Name & Class  \\    
\hline                        
       1 & Ell2       		&  Elliptical (5 Gyr old) \tablefoottext{a}  \\
       2 & Ell5   		&  Elliptical (2 Gyr old) \tablefoottext{a}  \\
       3 & Ell13         	&  Elliptical (13 Gyr old) \tablefoottext{a} \\
       4 & Arp220     		&  Starburst   		\tablefoottext{a}  \\
       5 & M82     		&      ''	\tablefoottext{a}  \\
       6 & IRAS 20551--4250   	&      ''	\tablefoottext{a}  \\
       7 & IRAS 22491--1808  	&      ''	\tablefoottext{a}  \\
       8 & NGC\,6240  		&      ''	\tablefoottext{a}  \\
       9 & S0     		&  S0          \tablefoottext{a}  \\
      10 & Sa     		&  Sa          \tablefoottext{a}  \\
      11 & Sb    		&  Sb          \tablefoottext{a}  \\
      12 & Sc    		&  Sc          \tablefoottext{a}  \\
      13 & Sdm  		&  Sdm         \tablefoottext{a}  \\
      14 & Sd   		&  Sd          \tablefoottext{a}  \\
      15 & Spi4  		&  Spiral      \tablefoottext{a}  \\
      16 & Sey18 		&  Seyfert 1.8    \tablefoottext{a}  \\
      17 & Sey2  		&  Seyfert 2      \tablefoottext{a}  \\
      18 & IRAS 19254-7245 	&  Seyfert 2       \tablefoottext{a}\\
      19 & QSO2   		&  QSO 2          \tablefoottext{a}   \\
      20 & hyb1\_gal10\_agn90 	&  Hybrid 10\% S0 + 90\% QS02        \tablefoottext{b} \\
      21 & hyb1\_gal20\_agn80 	&    ...          	   \\
      22 & hyb1\_gal30\_agn70 	&    ...          	   \\
      23 & hyb1\_gal40\_agn60 	&    ...          	   \\
      24 & hyb1\_gal50\_agn50 	&    ...          	   \\
      25 & hyb1\_gal60\_agn40 	&    ...          	   \\
      26 & hyb1\_gal70\_agn30 	&    ...          	   \\
      27 & hyb1\_gal80\_agn20 	&    ...          	   \\
      28 & hyb1\_gal90\_agn10	&  Hybrid 90\% S0 + 10\% QS02	   \\
      29 & hyb2\_gal10\_agn90	&  Hybryd 10\% I22491 + 90\% TQSO1	  \tablefoottext{b} \\
      30 & hyb2\_gal20\_agn80	&    ...          	   \\
      31 & hyb2\_gal30\_agn70	&    ...          	   \\
      32 & hyb2\_gal40\_agn60 	&    ...           	   \\
      33 & hyb2\_gal50\_agn50 	&    ...           	   \\
      34 & hyb2\_gal60\_agn40 	&    ...           	   \\
      35 & hyb2\_gal70\_agn30	&    ...            	  \\
      36 & hyb2\_gal80\_agn20	&    ...            	   \\
      37 & hyb2\_gal90\_agn10	&   Hybryd 90\% I22491 + 10\% TQSO1 	   \\
      38 & QSOL   		&    QSO  (low-luminosity)        \tablefoottext{b} \\
      39 & QSOH  		&    QSO  (high-luminosity)       \tablefoottext{b} \\
      40 & TQSO1 		&    QSO composite            \tablefoottext{a} \\
      41 & QSO1  		&       ''                \tablefoottext{a} \\
      42 & syth\_qso--0.25 	&    QSO synthetic\tablefoottext{c} \\
      43 & syth\_qso--0.50  	&       ''                 \\
      44 & syth\_qso--0.75	&       ``                 \\
      45 & syth\_qso--1.00 	&       ''                 \\
      46 & syth\_qso--1.25 	&       ``                 \\
      47 & QSO\_Cristiani	&    QSO           \tablefoottext{d} \\
      48 & QSO\_VVDS		&    QSO composite \tablefoottext{e} \\
      49 & QSO\_vandenBerk	&    QSO composite \tablefoottext{f} \\
      50 & Mrk231    		&    BALQSO        \tablefoottext{a} \\
\hline                             
\end{tabular}
\tablefoot{ \\
\tablefoottext{a}{Polletta et al. (2007).}
\tablefoottext{b}{Salvato et al. (2009).}
\tablefoottext{c}{\textit{LePhare} template database}
\tablefoottext{d}{Cristiani \& Vio (1990).}
\tablefoottext{e}{Gavignaud et al. (2006).}
\tablefoottext{f}{Vanden Berk et al. (2001).}
}
\label{Table_templates}
\end{table}
%

   \begin{figure}[!t]
   \centering
    \includegraphics[width=9.2cm,clip=true,trim=42 5 10 11]{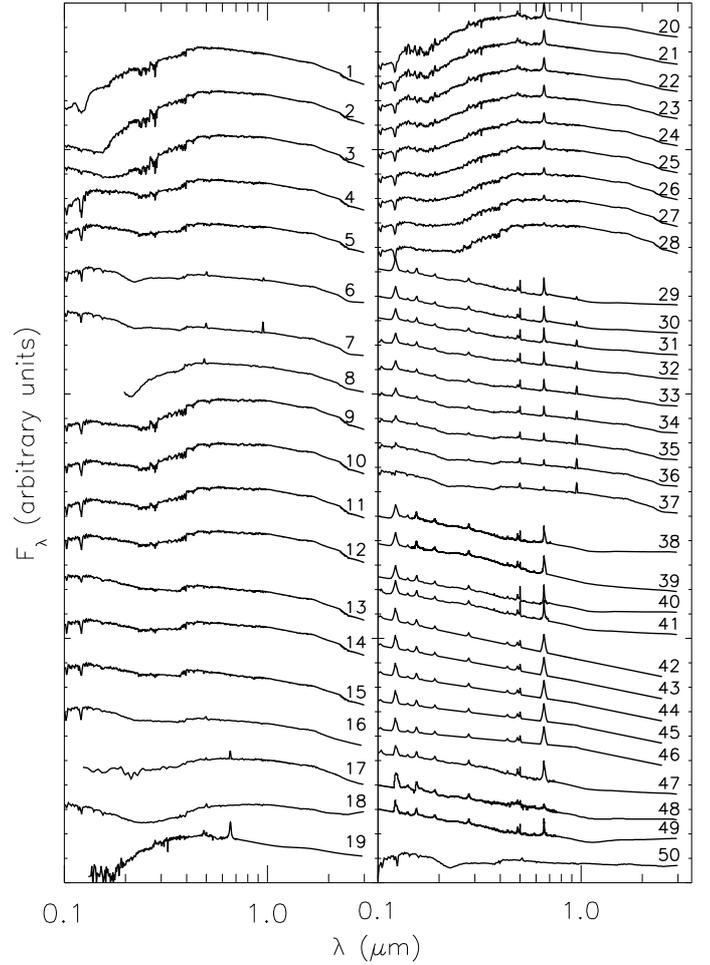}
      \caption{\scriptsize The complete extragalactic (galaxy + AGN) template
        database used in this work. The source and spectral class for each
        template given in Table \ref{templates} and described in Sect.\,3.1.}
     \label{templates}
   \end{figure}

\section{QSO photo-$z$ determination}
\label{section:QSO-photoZ-method}

We  used the publicly available template fitting code
\textit{LePhare}\footnote{http://www.oamp.fr/people/arnouts/LE\_PHARE.html}
(Arnouts et al. 1999, Ilbert et al. 2006) to estimate redshifts for our
selected QSOs. The code matches the photometric data of each ALHAMBRA QSO
source to a library of available templates providing the best-fit, spectral
classification, and photometric redshift by means of a $\chi ^2$ minimization
process. The minimization process accepts the inclusion of user-supplied
priors, different extinction laws, and the possibility to apply systematic
offsets to the different photometric bands in order to achieve the best match
between the colors of the sample and those provided by the template
library. The full capabilities and possibilities of the \textit{LePhare}
minimization code was extensively discussed by Ilbert et al. (2006,
2009).  Our final selection of templates, adopted reddening laws, 
priors, and systematic offsets are discussed in the following sections.

\subsection{Template selection}

The list of extragalactic templates used in this work are detailed in
Table\,\ref{Table_templates} and Figure\,\ref{templates}. The selection
includes SEDs for QSOs, Seyferts, starburst, normal galaxies, and stars. To
include low luminosity BLAGN that are partially or completely dominated by their
host galaxy light, we adopted the hybrid templates (consisting of a 
mixture of QSO and host galaxy SEDs) introduced by Salvato et al. (2009).
The variety of templates is justified by the need to test the ability of 
the ALHAMBRA survey to differentiate broad-line AGN emission from that of 
other extragalactic sources or stars, and enable us to do a blind search 
for these sources (Matute et al., in preparation).

In Tab.\,\ref{Table_templates} and Fig.\,\ref{templates}, the templates are
organized as:
\begin{itemize}

\item \textit{Non--active and starburst galaxies}: This includes 3 elliptical
  templates of different ages (2, 5, and 13 Gyr; \#1--3), the starburst
  galaxies Arp\,220, M\,82, NGC\,6240, IRAS\,20551, and
  IRAS\,22491\footnote{May contain an AGN responsible for 20\% of the
    bolometric flux (e.g. Veilleux et al. 2009)} (\#4--8) and 7 spirals (S0
  through Sd; \#9--15). They are all part of the SED library published by
  Polletta et al. (2007).

\item \textit{Obscured BLAGN}: Includes the Polletta et al. (2007) composite
  templates of a Seyfert\,1.8 and a Seyfert\,2 and the Seyfert--2 IRAS\,19254
  (\#16--18). The high-luminosity obscured sources are represented by a
  type--2 QSO (QSO2) and the BALQSO Mrk\,231 templates (Polletta et al. 2007;
  \#19 \& \#50). We also added the hybrid template library of Salvato et
  al. (2009) defined by 9 different combinations of a S0 and a QSO2 template
  (\#20--28).

\item \textit{QSO and hybrid--QSO templates}: Here we considered both the high
  and low luminosity SDSS composites (indices \#38, \#39), two templates from
  Polletta et al.  (2007; QSO1 and TQSO1 with indices \#40 and \#41), the
  Cristiani \& Vio QSO SED (\#47), the VVDS mean QSO SED (Gavignaud et
  al. 2006; \#48), and the mean QSO from Vanden Berk et al. (2001; \#49) also
  based on SDSS data. Hybrid templates include 9 different combinations of the
  starburst/ULIRG IRAS\,22491 template and a QSO1 template (Salvato et al. 2009; 
  \#29--\#37). As
  the quality and accuracy of the fit improved in several cases, we completed
  the list with a set of 5 synthetic QSO templates (\#42--46), covering
  continuum slopes ($\nu ^\alpha$) from $\alpha=-0.25$ to $\alpha=-1.25 $ (in
  steps of 0.25) below 1\,$\mu$m and fixed at $\alpha=-0.7$ above 1\,$\mu$m
  (\textit{LePhare} template database and references therein).

\end{itemize}

Finally, the stellar template database includes 131 spectra from the Pickels
(1998) stellar library plus 4 spectra of white dwarfs from Bohlin, Colina \&
Finlay (1995) and 19 additional templates from the \textit{LePhare} stellar
library.  Stellar templates were also included because white dwarfs and F/G stars
QSOs have similar colors to F/G stars in the $z$=[2--3] redshift
interval. Thus, our final database contains 204 templates (50 extragalactic +
154 stellar). During the minimization process, Galactic and extragalactic
templates were used separately.


\subsection{Extinction}


\begin{figure*}[t]
 \centering
 \includegraphics[width=16.6cm,angle=0,clip=true]{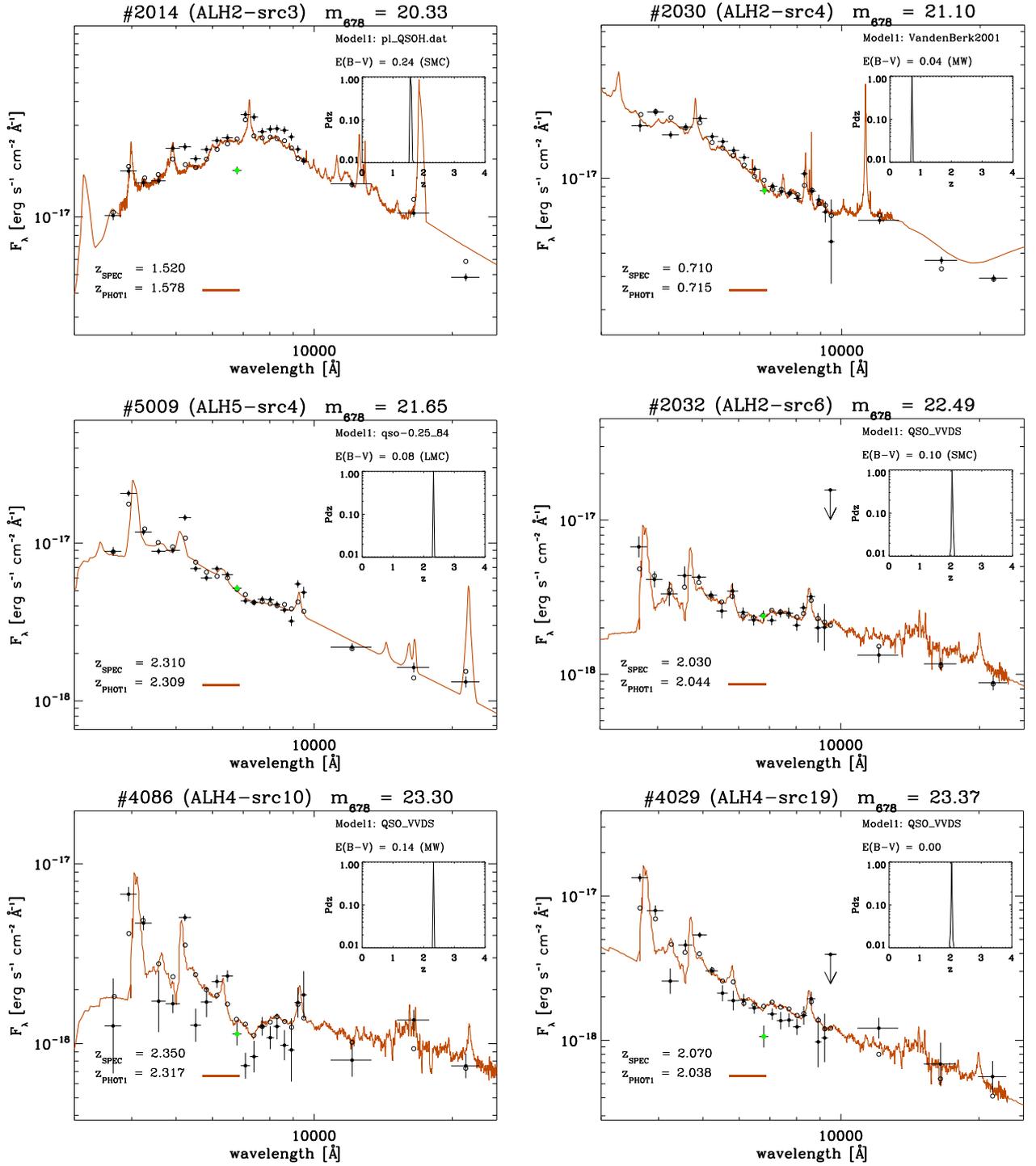}
\caption{\scriptsize Examples of best--fit solutions assuming a SMC extinction
  law for 6 sources covering a wide range of magnitudes
  ($\sim20$\,$\le\,m_{678}\le\,\sim23.5$) and spectroscopic redshifts ($0.7\le
  z\le2.3$).  Each panel includes the observed photometry, associated errors,
  and FWHM for each of the 23 ALHAMBRA filter set (black dots, vertical and
  horizontal error bar respectively). Photometric upper limits are indicated
  by arrows.  The continuous line shows the best-fit solution, while the open
  circles give the expected magnitude from the model corrected from systematic
  offsets. Additional info for each source includes: model name, reduced $\chi
  ^2$, amount of extinction, the normalized probability distribution as a
  function of $z$ (Pdz), the spectro--$z$ (and its source catalog), and the
  best--fit photo--$z$ solution. The title of each panel is labeled with the
  source ID in the ALHAMBRA catalog and the measured magnitude in the
  $m_{678}$ filter (green dot).}
\label{Fig:SEDfit_examples2c}
\end{figure*}

The photometry for several of the sources analyzed in this work defines a
continuum that strongly deviates from single or even double power-laws, 
most probably because of dust obscuration. For many of these types
of sources, there is an ongoing debate about whether (some of) these red QSOs are
obscured by either dust or an intrinsically red continuum (Richards et al. 2003; Young,
Elvis \& Risaliti 2008). Thus, our multiband template fit takes into
account the possibility of intrinsic dust obscuration within the source. We
adopted the Small Magellanic Cloud (SMC) extinction law (Prevot et
al. 1994), which has been shown to reproduce the observed reddening for
mildly obscured QSOs at $z<2.2$, for which there are no indications of the
Galactic feature at 2175\AA\, (Hopkins et al. 2004; Richards et al. 2003, York
et al. 2006). Gallerani et al. (2010) appeared to measure
some deviation from the SMC extinction law for higher redshift sources which is
one reason for adopting alternative extinction laws (see below). 
The attenuation due to dust ($A_V$) is given as a function of the
color excess $E(B-V)$ as $A_V=R_V\times E(B-V)$. We assumed $R_V=3.1$ and
a color excess in the range [0,1].

Furthermore, as our spectroscopic sample includes lower redshift Seyfert 1
nuclei, which may have a strong host-galaxy contribution, we considered
alternative extinction laws as the dust present in different galaxy types follow
extinction curves that deviate from that of the SMC. These deviations include
variations of the steepness in the attenuation curve as in the starburst
extinction law derived by Calzetti et al. (2000), or the presence of a broad
bump around 2175\AA\, as found for the Milky Way (MW; Seaton et al. 1979; Cardelli
et al. 1989) or the Large Magellanic Cloud (LMC; Fitzpatrick 1986).
Therefore, to reproduce  normal galaxy and starburst spectra
we also considered in the $\chi ^2$ minimization solutions based on the
LMC, MW, Calzetti's law, and Calzetti's law plus the absorption feature around
2175\,\AA.  In this case, the minimization process takes into account all
possible SEDs and extinction laws simultaneously, choosing the best suited to
each source. These additional extinction laws allow us to probe their
relevance to the accuracy of the results.

The light attenuation by the inter-galactic medium (IGM) was taken into
account internally by \textit{LePhare} following the opacity curves, binned into
redshift intervals of $\Delta z = 0.1$, published by Madau (1995).

\subsection{Systematic offsets}

Photometric redshifts depend strongly on the precision of the photometry
and the capabilities of the template database to reproduce the colors of the
source population as a function of $z$. If we were to assume that the selected
template database is representative of our source population, then for a given
filter the average deviation between the observed flux and the best-fit
predicted flux should be zero for normally distributed uncertainties. If this
is not the case, a zero-point offset must be applied to the photometry derived
from the template database when there is a non-zero average deviation
between the observed and predicted fluxes. The ALHAMBRA photometric
calibration is based on a selection of NGSL stars following the
methodology discussed in Aparicio--Villegas et al. \cite{Aparicio2010}. We note here that
we have not modified this criteria. The computed offsets instead, ''adapt`` the
templates to provide a better fit the observed photometry. Therefore, alternative
templates or object selection can (and will) lead to different offsets (e.g. Table 1 by
Ilbert et al. 2006).

We used the spectroscopic sample described in Sect,\,2 and the
colors in the filters A457M, A646M, and A829M to compute these systematic
offsets using an iterative approach of find the best-fit SED for each source,
deriving the mean deviations for each filter, applying offsets, re-computing best-fit
SEDs, etc. The iteration process was halted when the variation in $\chi^2$ between
iterations drops below 2\%. In general, the procedure did not require more than 4
iterations to converge. We found that the offsets to be applied are small
and agree with the typical photometric error for each band in the
sample.  Table \ref{table:ALHfilters} reports the values of these corrections
as well as the typical photometric error for each band in columns 5 and 7,
respectively.

   \begin{figure}[t]
   \centering

    \hspace{-0.5cm}
    \includegraphics[width=6.8cm,angle=-90]{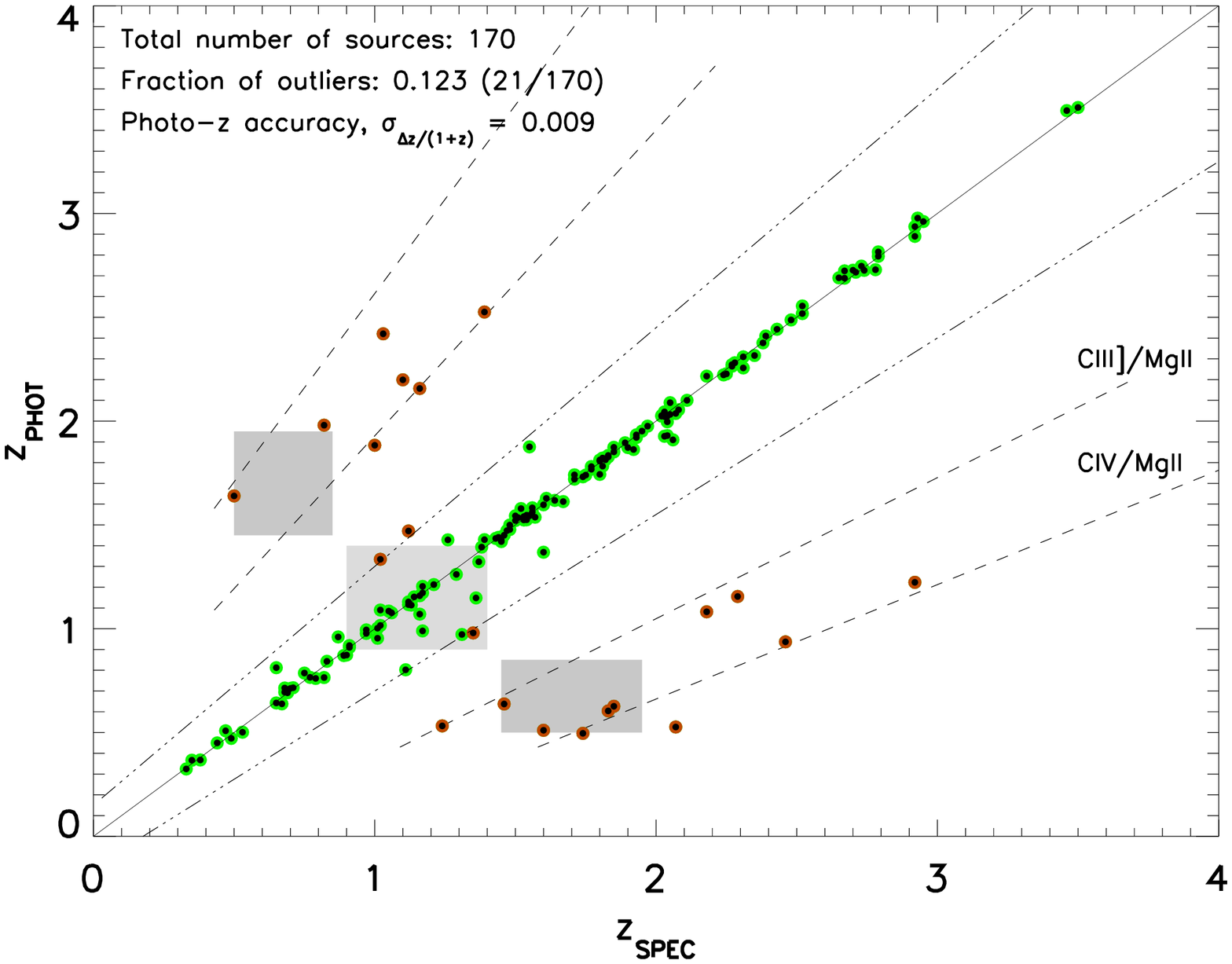}
    \includegraphics[width=7.5cm]{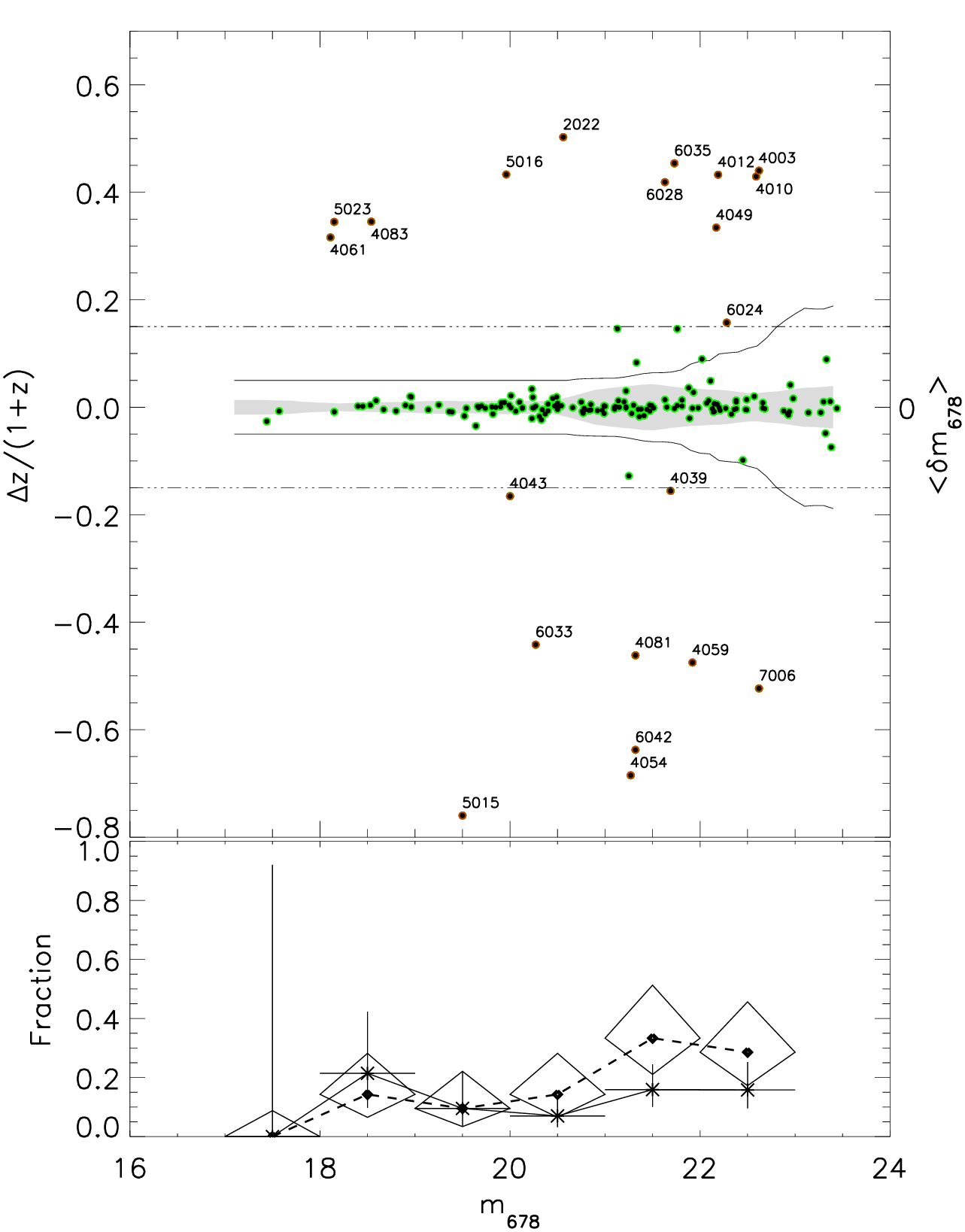} 
      \caption{\scriptsize Photo-$z$ efficiency using several extinction laws
        (the MEL solution). \textit{Top}: Comparison between the best fit
        photo-$z$ solution and the measured spectro-$z$ shows a good agreement
        between the both.  The continuous line gives the $z_{phot} = z_{spec}$
        relation while the dashed line represent the boundary between good
        solutions (green dots) and outliers (red, indexed dots) and defined as
        $|\Delta z| /(1+z_{\mathrm{spec}}) > 0.15$. \textit{Central}:
        Distribution of $\Delta z /(1+z_{\mathrm{spec}})$ as a function of the
        ALHAMBRA magnitude $m_{678}$. The mean magnitude error, per magnitude
        interval of $\Delta m=1$, of the filter A678M is indicated by the
        continuous lines. The accuracy per magnitude interval ($\Delta m = 1$)
        is highlighted by a grey shaded area.  This accuracy shows a small
        correlation with apparent magnitude. \textit{Bottom:} This panel shows the contribution
        of each magnitude bin ($\Delta m=1$) to the outlier population as
        filled diamonds connected by a dashed line.  The magnitude intervals
        considered and the associated errors are indicated by the large
        diamonds. The fraction of outliers with respect to the number of sources in
        the same magnitude intervals are given by asterisks connected by a
        continuous line. The magnitude intervals and the associated errors are
        indicated by the large vertical and horizontal lines. In both cases,
        errors are assumed to be Poissonian and were calculated following
        Gehrels (1986). 
        }
    
     \label{Fig:SMC-accuracy2}
   \end{figure}
%

\subsection{Priors}

The introduction of important \textit{a priori} information into the redshift
probability distribution function (Pdz) based on Bayesian probability can in
many cases improve the quality of the solutions by favoring a particular
redshift based on known redshifts and/or color distributions (e.g. Ben\'itez
2000). Our analysis only makes use of a particular luminosity and redshift range
prior and does not include any redshift distribution or color information of
known BLAGN/QSO populations. We restricted the permitted absolute magnitudes
in the A457M ($\lambda 4575\AA$) filter between -17 and -28. Absolute
magnitudes in this filter are consistent with the commonly used broad-band
standard filter $B$. This includes not only the typical range where most
BLAGN/QSOs are found ($-28\le M_{457} \le -20$) but also the range for
host-dominated BLAGN and normal galaxies (Salvato et al. 2009; Polletta et
al. 2007; Rowan-Robinson et al. 2008).

\subsection{Photo-$z$ determination summary}

We used the code \textit{LePhare} to estimate photometric redshifts for 170
spectroscopically identified BLAGN and QSOs with high quality ALHAMBRA
photometry. For the $\chi^2$ minimization process, we considered the
following:

\begin{itemize}
  \item A database of 204 templates: 154 stellar and 50 extragalactic.
  \item Several extinction laws: MW, LMC, SMC, and Calzetti's starburst laws with a
    color-excess range of $E(B-V)$=[0.0,1.0].
  \item Given our template library, we made a correction to the zero-point of the
    filters that show a non-zero average deviation between the observed and
    best-fit predicted magnitudes.
  \item A simple luminosity prior of -17$\,\le M_{A457M}\le\,$-28.
  \item A redshift space interval of $0\le z\le6$ binned in redshifts intervals of $\delta$z=0.04.
\end{itemize}

Figure\,\ref{Fig:SEDfit_examples2c} shows an example of the excellent
agreement between the data and the fitted template for 6 sources with a wide
range of magnitudes ($20\le m_{A678M}\le 23.5$), redshifts ($0.7\le z\le2.3$),
and intrinsic extinctions ($0.0\le$ $E[B-V]\le0.2$).

\section{Results and discussion}
\label{section:Results}

\subsection{Photo-$z$ accuracy}
\label{subsection:photoz-accuracy}

The efficiency of the photo-$z$ determination is quantified by comparing
the spectroscopic redshifts (hereafter spectro-$z$) of
170 sources in our BLAGN/QSO sample. Photometric redshifts are generally
characterized by both their accuracy and outlier fraction. The accuracy is
defined as the standard deviation of $\Delta z/(1+z_{\mathrm{spec}})$, denoted
$\sigma_{\Delta z/(1+z_{\mathrm{spec}})}$, where $\Delta z =
z_{\mathrm{spec}}-z_{\mathrm{phot}}$, while the outlier fraction ($\eta$) is
defined as the fraction of sources with catastrophic solutions (i.e.
solutions that are inconsistent with the measured spectro-$z$). 
In our analysis, we assumed that a source is an outlier if $|\Delta z|/(1+z) \ge 0.15$. 
This value was selected \textit{a priori} to be compatible with the cutoff of similar
studies (e.g. Luo et al. 2010, Salvato et al. 2009, Ilbert et al. 2009,
Rowan-Robinson et al. 2008). An alternative accuracy estimate that has been used
by several authors (e.g. Ilbert et al. 2006, Brammer et al. 2008) is the
normalized median absolute deviation (NMAD) defined as

\begin{center}
$\sigma_{\mathrm{NMAD}}=1.48\times \mathrm{median}|\frac{\Delta z - \mathrm{median}(\Delta z)}{1+z_{\mathrm{spec}}} |$.
\end{center}
  \begin{figure}[t]
   \centering
    \includegraphics[width=6.5cm,angle=-90]{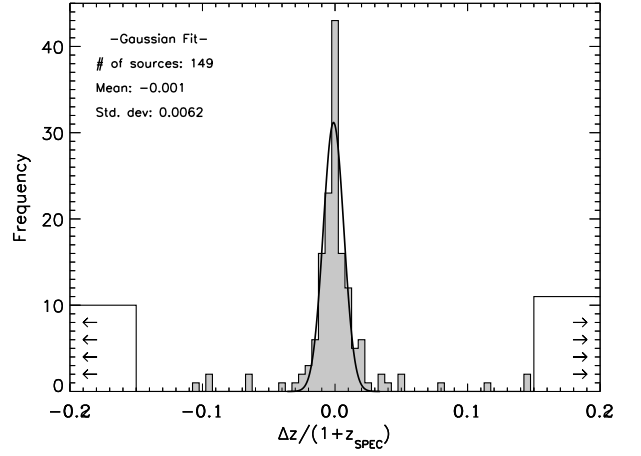}
    \caption{\scriptsize Uncertainty distribution, $\Delta z /(1+z)$, for the
      170 BLAGN/QSOs in our sample considering the MEL approach.  Non-outlier
      and outliers are represented by filled and open histograms,
      respectively. The continuous line represents the best Gaussian fit to the
      observed distribution of non-outliers. The number of non-outliers (\#),
      the center, and $\sigma$ of the best-fit Gaussian distribution are
      indicated.}
     \label{Plot:Outlier_selection}
   \end{figure}

This parameter can be directly compared to the standard deviation of $\Delta
z/(1+z_{\mathrm{spec}})$ in the case of normal distributions and has the
advantage of being less sensitive to outliers. From now on, we use
$\sigma_{\mathrm{NMAD}}$ as our estimate of the photo-$z$ accuracy.

We now discuss our results based on the number of extinction laws considered in
the computation, namely either a single (SMC) extinction law (SEL hereafter) or multiple
(SMC, LMC, Milky--Way, and Calzetti) extinction laws (MEL hereafter).  Table
\ref{table:Fit_results} describes the solutions found for the two sets of
extinction laws considered. In the case of a SMC extinction law, we obtained an
accuracy of $\sigma_{\mathrm{NMAD}}=0.009$ with a fraction of outliers of
$\eta \sim 12$\% (21 out of 170 sources).  A comparison between the derived
photo-$z$ and the spectro-$z$ is shown in the top panel of
Fig.\,\ref{Fig:SMC-accuracy2}.  The narrow scatter present for the good fits
(green dots) is highlighted by the distribution of the source redshift
accuracy that lies in the range $|\Delta z|/(1+z_{\mathrm{spec}})\leq 0.15$ (i.e. no
outlier region) and shown in Fig.\,\ref{Plot:Outlier_selection}. This
distribution is well-represented by a Gaussian with no measurable bias
(centered at $-0.001$) and a $\sigma$ of $\sim$0.006. Identical results were
found when we considered several extinction laws during the minimization process
($\sigma_{\mathrm{NMAD}}=0.009$, $\eta \sim 12$\%) but, as we see in the
following paragraphs, the MEL approach is able to more accurately reproduce the SED
distribution of the BLAGN/QSO population.

Besides the ability to provide precise photo-$z$'s, our analysis allows us to
recover the correct SED for most of the sources.
Figure\,\ref{Fig:template-distrib} presents the distribution of the templates
for the MEL best-fit solutions. We did not find sources with stellar templates
that have best-fit solutions (i.e.  $\chi ^2 _{stellar}< \chi ^2 _{gal-QSO}$) and the
majority of the sources are best-fitted by pure type-1 QSO templates or QSO
hybrid templates (ULIRG/QSO1 template indices \#29 and above; Sect.\,3.1). When we
did not take into account the outlier fraction of the sources, we found that
95.3\% of them (142/149) have either a QSO or hybrid--QSO best-fit template, 1.3\% (2/149)
are fitted by a QSO2 or hybrid-QSO2 template, and 3.4\% (5/149) are fitted
with a normal or starburst template. Five sources have best-fit solution
templates compatible with a non-active SED. A closer look at the ALHAMBRA
photometry, the best fit solution, and the observed spectra (when available)
revealed that: $i)$ One source might be incorrectly classified as BLAGN since
both the spectra and the ALHAMBRA photometry point to an early-type galaxy;
$ii)$ one source belongs to the fainter part of the population
($m_{678}=23.33$) and the associated errors ($\Delta m >0.2$) have diluted any
possible BLAGN signature in the ALHAMBRA photometry; $iii)$ the other 3
sources have optical spectra compatible with galaxy templates with different
degrees of starforming and post-starforming activities, i.e. no signs of broad
emission and a continuum with a well-defined 4000\,\AA\, break, but with
ALHAMBRA photometry showing some degree of AGN activity (blue continuum and
indications of faint, typical BLAGN line-emission namely of \mgii, \ciii or \civ at
the spectroscopic redshift of the source). The incorrect solutions found for these 3
sources are probably a consequence of the uncertainties expected from the
method, particularly regarding the chosen template database and the absence of any 
variability correction of the photometry.  As a test of the degeneracy introduced by the
chosen template database, we considered an alternative database of only
QSOs and hybrid QSO/ULIRG templates (see indices \#29--39 in
Table\,\ref{table:templates} and Fig.\,\ref{templates}) for the 5 sources with
a normal galaxy or starburst best-fit solution. We found that: $i$) for 2
sources we were unable to recover in this case the photo-$z$, casting some
doubts on the spectral classification of the sources or an incomplete template
database; $ii$) a correct photo-$z$ was found for the other 3 sources where
hybrid templates with a weaker BLAGN component (10--20\%) were selected by the
best-fit solution. This last case highlights the degeneracy introduced by the
selected template database for certain host and BLAGN luminosity
regimes.  For this small fraction of sources (2\%; 3/149), the incorrect
spectral classification, using the method described here, will be taken into
account in any statistical analysis of the BLAGN/QSO population detected in
the ALHAMBRA fields to be presented in a forthcoming paper (Matute et al., in
preparation).

Furthermore, we note that the photo-$z$ determination described here is able
to recover the redshift of the sources in the interval $2.2\leq z \leq 3.6$,
which has been traditionally biased against the selection of QSOs because of their similar
colors to F/G stars. Hence, the photometry and the method described here
could provide an efficient way of both classifing and deriving a reliable photometric
redshifts for BLAGN/QSO candidates pre-selected, for example, by their X--ray
flux.  The catalog and derived luminosity functions for BLAGN/QSO selected
purely based on the ALHAMBRA photometry will be presented in a companion paper
(Matute et al. in preparation).

The dependence of our results on redshift, apparent magnitude of the source, and
the systematic offsets applied during the photo-$z$ computation
are described in the following sections. These dependences are similar in the
two sets of extinction laws unless otherwise specified.

%
\begin{table*}
\caption{BLAGN/QSO photometric redshift results}          
\label{table:Fit_results}     
\centering                         
\begin{tabular}{c c c c c c c c c}       
\hline\hline                 
& \multicolumn{4}{c}{Medium-bands + \textit{JHK$_S$}} & \multicolumn{4}{c}{Medium-bands} \\
& \multicolumn{2}{c}{SEL\tablefoottext{a}} & \multicolumn{2}{c}{MEL\tablefoottext{b}} & \multicolumn{2}{c}{SEL\tablefoottext{a}} & 
  \multicolumn{2}{c}{MEL\tablefoottext{b}} \\
  &   No offsets & Offsets & No offsets & Offsets & No offsets & Offsets & No offsets & Offsets \\ 
  & (1) & (2) & (3) & (4) & (5) & (6) & (7) & (8)\\
\hline 
$\sigma_{NMAD}$                          & 0.010 & 0.009 & 0.011 &  0.009  & 0.013  & 0.013 & 0.014 & 0.011 \\ 
$\eta$                                   & 15.9  & 12.3  & 14.1  &  12.3   & 16.4   & 15.3  & 19.4  & 15.3 \\ 
Extincted fraction \tablefoottext{c}     & 46.9  & 61.7  & 61.6  &  79.0   & 45.8   & 55.6  & 71.0  & 81.3 \\
$\langle E(B-V)\rangle$\tablefoottext{d} & 0.045 & 0.058 & 0.085 &  0.092  & 0.037  & 0.049 & 0.063 & 0.085 \\
QSO1--Hybrid fraction\tablefoottext{e}   & 92.3  & 92.6  & 95.2  &  95.3   & 92.2   & 91.7  & 94.9  & 95.1 \\                             
\hline                        
\hline                             
\end{tabular}
\tablefoot{\scriptsize \\
\tablefoottext{a}{Results using the SMC extinction law.}
\tablefoottext{b}{Results using several extinction laws (SMC, LMC, MW, and Calzetti).}
\tablefoottext{c}{Fraction of sources requiring extinction.}
\tablefoottext{d}{Mean color excess applied to the extincted sources.}
\tablefoottext{e}{Fraction of sources with pure QSO1 or hybrid QSO1/IRAS\,22491 templates.}
}
\end{table*}

\subsubsection{Dependence on redshift}
\label{sec:Dependency_with_z}

The accuracy of the photo-$z$ results is rather independent of the redshift
with the exception of the interval $z$=[0.9,1.4] (light--grey square in
Fig.\,\ref{Fig:SMC-accuracy2}a for the SMC results). The presence of only one
prominent line (\mgii) in this interval introduces some aliasing that depends on
the intensity of the line and how well is sampled by the ALHAMBRA
filters. This small degradation of the solution occurs when the peak of
the \mgii line falls within two ALHAMBRA filters.  The distribution of the
outlier fraction of the sources (Fig.\,\ref{Fig:SMC-accuracy2}, red dots)
follows a bimodal behaviour around $z_{\mathrm{spec}}\!\sim\!$ 1.4. 
Below this redshift ($z_{\mathrm{spec}}\!\leq\!$ 1.4),
the minimization process tends to overestimate the photo-$z$ solutions, while it
underestimates them at higher redshifts ($z_{\mathrm{spec}}\!>\!$ 1.4). This
effect can be explained by $i)$ the QSO color/redshift degeneracy (i.e. the
degree of similarity between the colors at different redshifts; e.g. Richards
et al. 2001) and $ii)$ a line misidentification (Croom et al. 2004). These
degeneracies, still present in the ALHAMBRA data but to a much lesser extent
than for broadband photometry, are highlighted as grey shaded areas and
dot--dashed lines in Fig.\,\ref{Fig:SMC-accuracy2} for the color--color and
line misidentification degeneracies. Further details of the origin of these
degeneracies are given in Sect.\,\ref{subsection: outlier nature} where we
explore the nature of the outlier fraction of sources.

\subsubsection{Dependency with apparent magnitude}
\label{sec:Dependency_with_m}

As highlighted in the central panel of Fig.\,\ref{Fig:SMC-accuracy2}, we do
not find any dependence of the accuracy on the apparent magnitude of the
source but only a clear degradation of the solutions is found at fainter magnitudes
($m_{678}\geq22.0$) caused by the slightly noisier photometry ($\Delta m_{678}
\sim 0.2$ at $m_{678}=23$ as indicated by the continuous line in the central
panel of Fig.\,\ref{Fig:SMC-accuracy2}). On the other hand, the outlier
fraction shows a moderate correlation with apparent magnitude (bottom panel of
Fig.\, \ref{Fig:SMC-accuracy2}), where $\sim$62\% of the outliers have
$m_{678}\geq21.0$. Nevertheless, although some outliers might be produced by 
noisier photometry, other factors might also contribute to the catastrophic 
failures (see Sect.\,4.3).


  \begin{figure}[!t]
  \centering
    \includegraphics[width=6.5cm,angle=-90,trim=0 0 10 0]{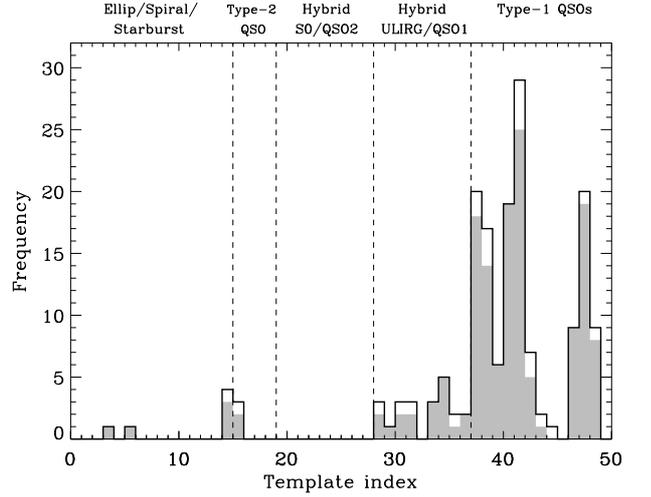}
      \caption{\scriptsize Spectral energy distribution for the MEL best--fit solutions.
	Open histogram takes into account all sources, while the shaded histograms
	consider only the sources with good photo-$z$ solutions. We find that none of
	the ALHAMBRA BLAGN/QSOs is well-represented by a stellar template. Of all the
	extragalactic templates considered, the majority (95.3\%; 142 out of 149) of the
	sources with good solutions (i.e. no outliers) have best-fit templates compatible
	with a pure QSO or hybrid QSO/ULIRG template.}   
     \label{Fig:template-distrib}
   \end{figure}
%


\subsubsection{Dependence on photometric offsets}
\label{sec:Dependency_with_Offsets}

Table \ref{table:Fit_results} reports the improvement in the fit achieved by
including the photometric offsets derived from the ALHAMBRA photometry and 
the template dataset. Although the accuracy is only marginally higher, 
we found that the greatest benefit is the
reduction in the numbers of outliers (from $\sim$16\% to $\sim$12\%). The same
behaviour has been previously observed by other authors (e.g. Ilbert et
al. 2009).

\subsection{Extinction distribution}
\label{sec:Dependency_with_extlaws}
The distribution of the color excess $E(B-V)$ required by the best-fit
solutions is shown in Fig.\,\ref{Fig:amount-extintion}. Considering only the
SMC extinction law (Fig.\,\ref{Fig:amount-extintion}, left), we found that
$\sim62\%$ (92 out of 149) of the non-outlier sources require some dust
extinction in their best-fit solution. The mean extinction of the sample is
$\langle E(B-V)\rangle \sim0.06$, with the majority of these sources
($\sim$53\%) having a color excess compatible with the mean Galactic value
along the line of sight ($E(B-V)\le0.05$).  Only 5 sources require extinctions
in the range $0.2\!\leq E(B-V)\!\leq0.4$. The solutions obtained using the
MEL approach required a slightly larger fraction of extincted sources (79\%)
with a mean color excess of $\langle E(B-V)\rangle \sim0.09$.  A significant
fraction of sources ($\sim$41\%) with small extinctions ($E(B-V)<0.05$) was
also found in this case. We note here that, since several
templates in our database are already extincted, the derived values of
$E(B-V)$ should only be considered as lower limits when these template are
chosen by the best-fit solution.

The presence of dust-extincted QSOs is no surprise as the SDSS survey has
established the existence of a non-negligible fraction of QSOs with spectral
indices ($f_{\nu} \propto \nu^{\alpha}$) redder than $\alpha=-1$ with
extinctions as high as $E(B-V)\sim 0.5$ (Gregg et al. 2002; Richards et
al. 2003). The extinction values found in this work for BLAGN/QSO are in good
agreement with the extinction interval derived by Richards et al.  (2003;
$E(B-V)$=[0.07--0.135]) in order to reproduce the spectral indices of red QSOs
using templates of QSOs with normal colors from the SDSS. A detailed analysis
of the selection effects introduced by intrinsic dust extinction will be
carried out for the future catalog release of QSOs detected by the ALHAMBRA
survey (Matute et al., in preparation).

\subsection{Nature of the outliers}
\label{subsection: outlier nature}

  \begin{figure*}[ht]
  \centering
    \vspace{0.7cm}
    \includegraphics[width=6cm,angle=-90]{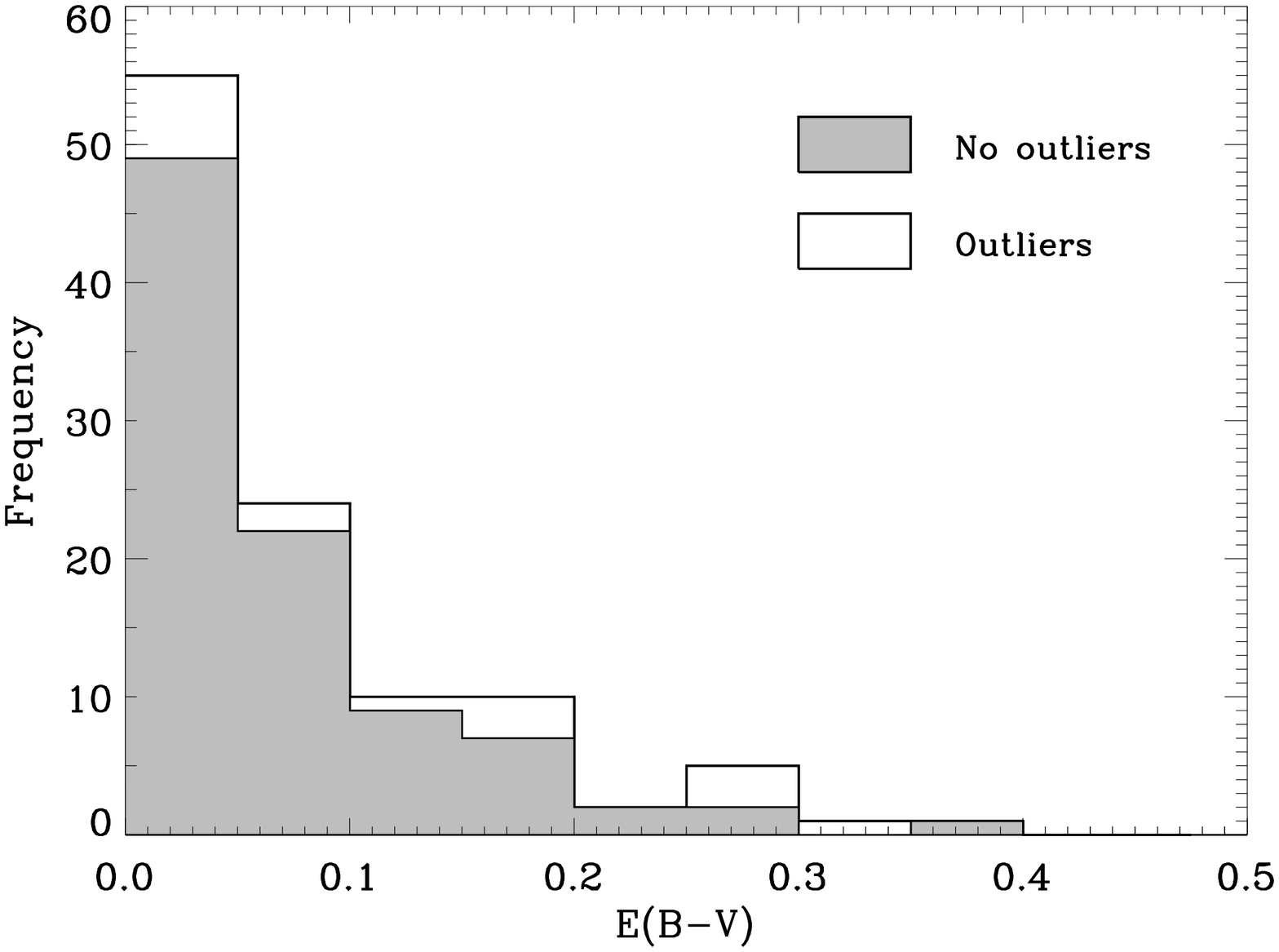}
    \includegraphics[width=6cm,angle=-90]{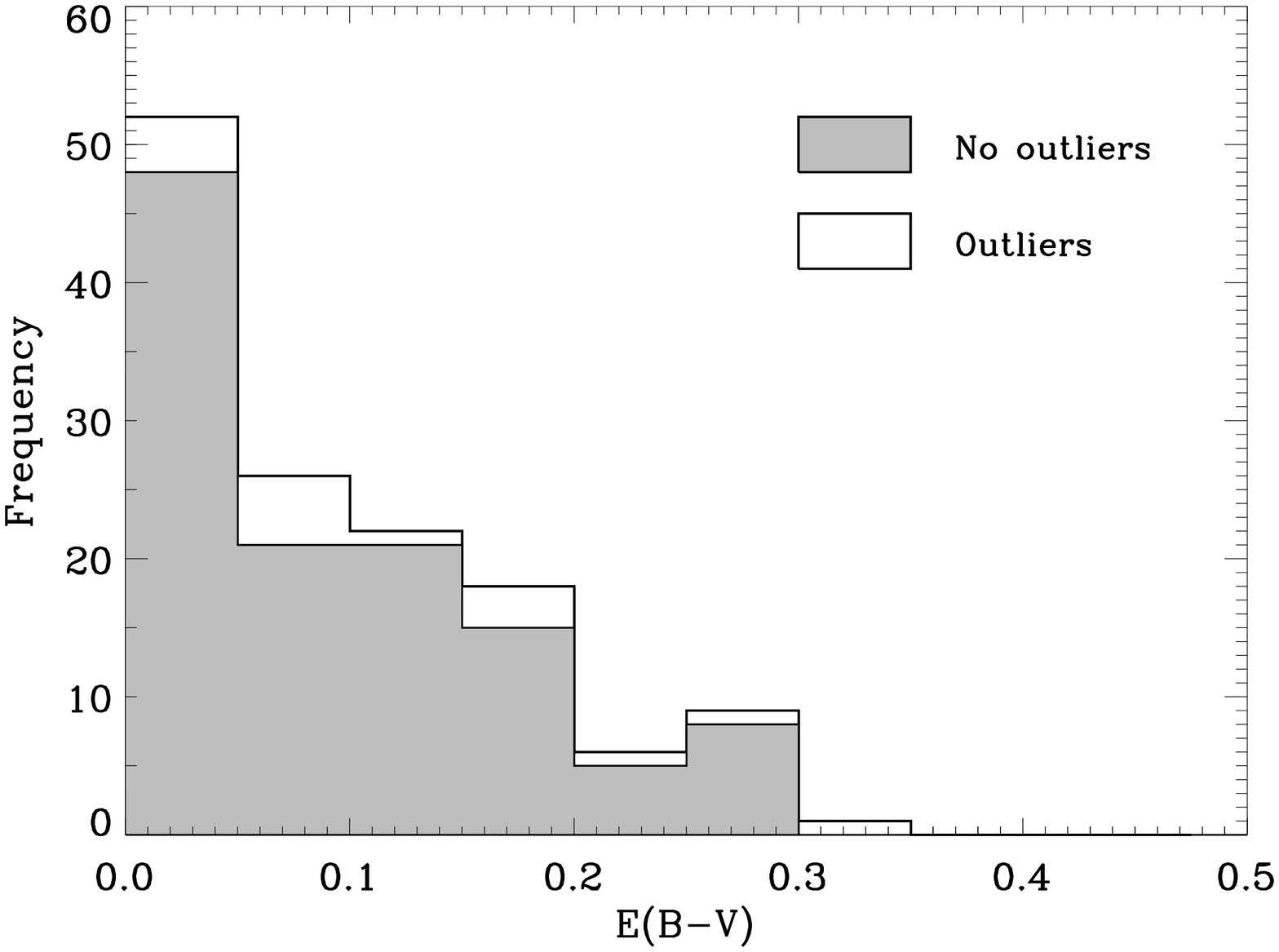}
     \vspace{0.5cm}
      \caption{\scriptsize Color excess distribution for the sources that required a certain 
	amount of extinction, E(B-V)$>$0, to be applied to the 
	best--fit template. \textit{Left}) Color excess distribution using only the SMC 
	extinction law. \textit{Right}) Color excess distribution using various 
	extinction laws (see Sect.\,3.2). Histograms as in Fig.\,\ref{Fig:template-distrib}.}   
     \label{Fig:amount-extintion}
   \end{figure*}
%


To fully characterize the photo-$z$ estimate for the ALHAMBRA QSOs,
it is fundamental to understand the reasons for an incorrect photo-$z$
determination. In the following, we describe the photometry and nature of
the $\sim$12\% of the sources (21/170) with inaccurate photo-$z$
determinations in an attempt to infer the reasons for the incorrect
solutions. The outlier best-fit templates (Fig.\,\ref{Fig:template-distrib},
open fraction of the histogram), extinction values
(Fig.\,\ref{Fig:amount-extintion}, open fraction of the histogram), and
redshift distribution (upper panel, Fig.\,\ref{Fig:SMC-accuracy2}) are similar
to those of the remaining population. The most probable reasons for the
catastrophic failures are enumerated in the following paragraphs. We only
discuss the MEL fit approach as it offers the best results, but similar
conclusions can be drawn from the SEL approach.

1) \textit{The less precise photometry of fainter sources}. The apparent
dependence of the outlier fraction on magnitude has already been
discussed in Sect. 4.1.2. This dependence is almost expected as the larger
photometric errors of these fainter objects (continuous line in the central
panel of Fig.\,\ref{Fig:SMC-accuracy2}) broadens the Pd$z$ distribution,
increasing the number of peaks and directly influencing the photo-$z$
results. Nevertheless, only 2 outliers ($\sim$10\% of all the outliers;
Fig.\,\ref{Fig:SMC-accuracy2} \#'s 4049 \& 6024) have large enough photometric
uncertainties ($\Delta m > 0.2$) that can explain their incorrect photo-$z$
solution.

2) \textit{Emission-line misclassification}. Eight outlier solutions (38\% of
all the outliers; Fig.\,\ref{Fig:SMC-accuracy2} \#'s 2022, 4010, 4061, 5015,
5023, 5016, 6028 and 6035) misclassify either a single or pair of emission
lines depending on the redshift of the source and on how intense and well-sampled 
the lines are by the ALHAMBRA photometry (see Sect.\,\ref{sec:Dependency_with_z}). 
The expected source distribution for single
line misclassifications (confusion of the \ciii and \civ lines with \mgii and
viceversa) is outlined in the top panel of Fig.\,\ref{Fig:SMC-accuracy2} by
the dashed lines. The emission-line pairs \mgii-- \hbeta\,\, and \ciii-- \mgii
can be confused in the redshift intervals z$\sim$[0.6--0.9] and
z$\sim$[1.5--1.9] owing to the limited resolution of the ALHAMBRA spectra
(i.e. the ability to resolve the line given its intensity and position with
respect to the ALHAMBRA filters). This latter source of confusion is
highlighted as the dark grey areas in the top panel of
Fig\,\ref{Fig:SMC-accuracy2}. For 6 out of the 8 sources, we found secondary
solutions in agreement with the spectro-$z$. For these sources, additional
\textit{a priori} information would be needed in order to favor one solution over
another. A simple \textit{a posteriori} condition that weights the solution
at each redshift (Pdz), given the source magnitude, was derived from
known luminosity functions of optically selected QSOs. We found that this
approach does not always reduce the fraction of outliers in our sample while,
if applied to the whole ALHAMBRA survey, may introduce a bias. 

  \begin{figure}[!b]
  \centering
    \vspace{0.7cm}
    \includegraphics[width=7cm,angle=0,trim=0 0 10 0]{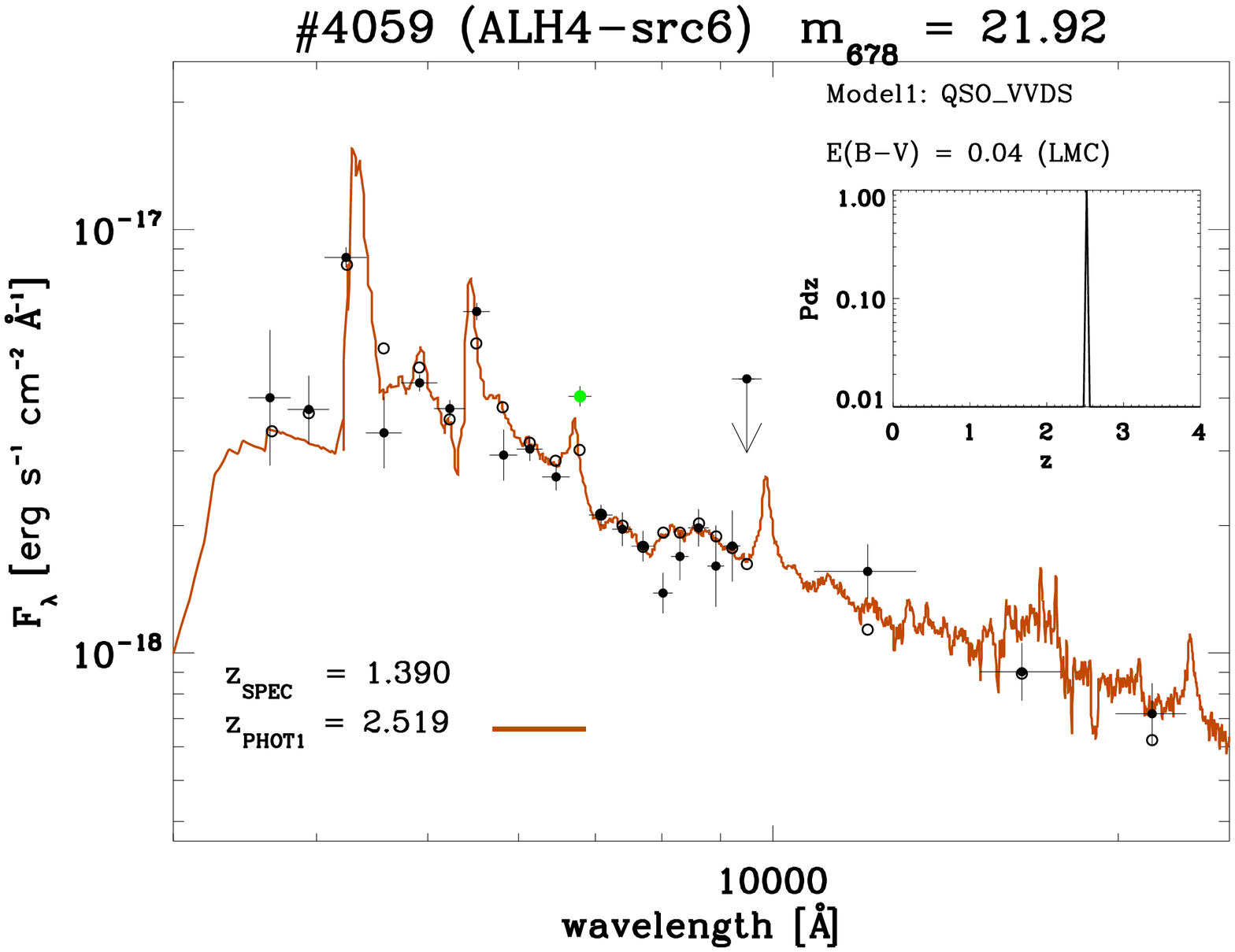}
     \vspace{0.9cm}
      \caption{\scriptsize Example of a source with a dubious spectro-$z$ of
        1.39. The ALHAMBRA photometry and errors, represented by the black
        dots and crosses, provide a photo-$z$ solution (continuous line) of
        2.53 with peak emission-lines from L$_\alpha$, \ciii, and \civ clearly
        visible at $\sim$4300, $\sim$5400 and $\sim$6700\,\AA \, respectively.
        Lines and symbols as in Fig.\,\ref{Fig:SEDfit_examples2c}. The spectra
        of this source is currently unavailable to the public.}   
     \label{Fig:wrong_zspec}
   \end{figure}


3) \textit{The possibility of an incorrectly assigned spectro-$z$}. The
ALHAMBRA photometry and the solutions found for 3 sources ($\sim$14\% of all
the outliers; Fig.\,\ref{Fig:SMC-accuracy2} source \#'s 4059, 4081 and 7006) seem to
indicate that an incorrect spectroscopic redshift has been assigned. An
example of one of these sources is shown in Fig.\,\ref{Fig:wrong_zspec}, where
the ALHAMBRA low-resolution spectra show a well-defined power-law continuum
with signs of two or more intense emission-lines that do not correspond to
the assigned spectro-$z$. A low S/N spectra and/or limited wavelength coverage
are typical causes of an incorrect assignment of spectroscopic
redshifts (Fern\'andez--Soto et al. 2001). Two out of three sources are
located in the COSMOS field. These sources spectro-$z$ come from VIMOS/VLT
observations (Brusa et al. 2010) that are not included in the public release of the
zCOSMOS database. Some degeneracy in the spectro-$z$ is expected given the
limited wavelength range covered by the VLT/VIMOS observations of
$\lambda\sim\,[5500, 9500]\,\AA$. This means that just a single broad emission-line 
will be visible in certain redshift intervals (e.g. confusion may arise
between \mgii at $z\sim[0.8,1.5]$ and C$_{\mathrm{III}}$ at $z\sim[2.2,2.6]$),
as seems the case for these 3 outliers. Although further confirmation of their
true redshifts is required, these sources prove how precise photo-$z$ can
complement spectro-$z$ making them an invaluable tool in modern observational
cosmology.

  \begin{figure}[!b]
  \centering
    \vspace{0.7cm}
    \includegraphics[width=7cm,angle=0,trim=0 0 10 0]{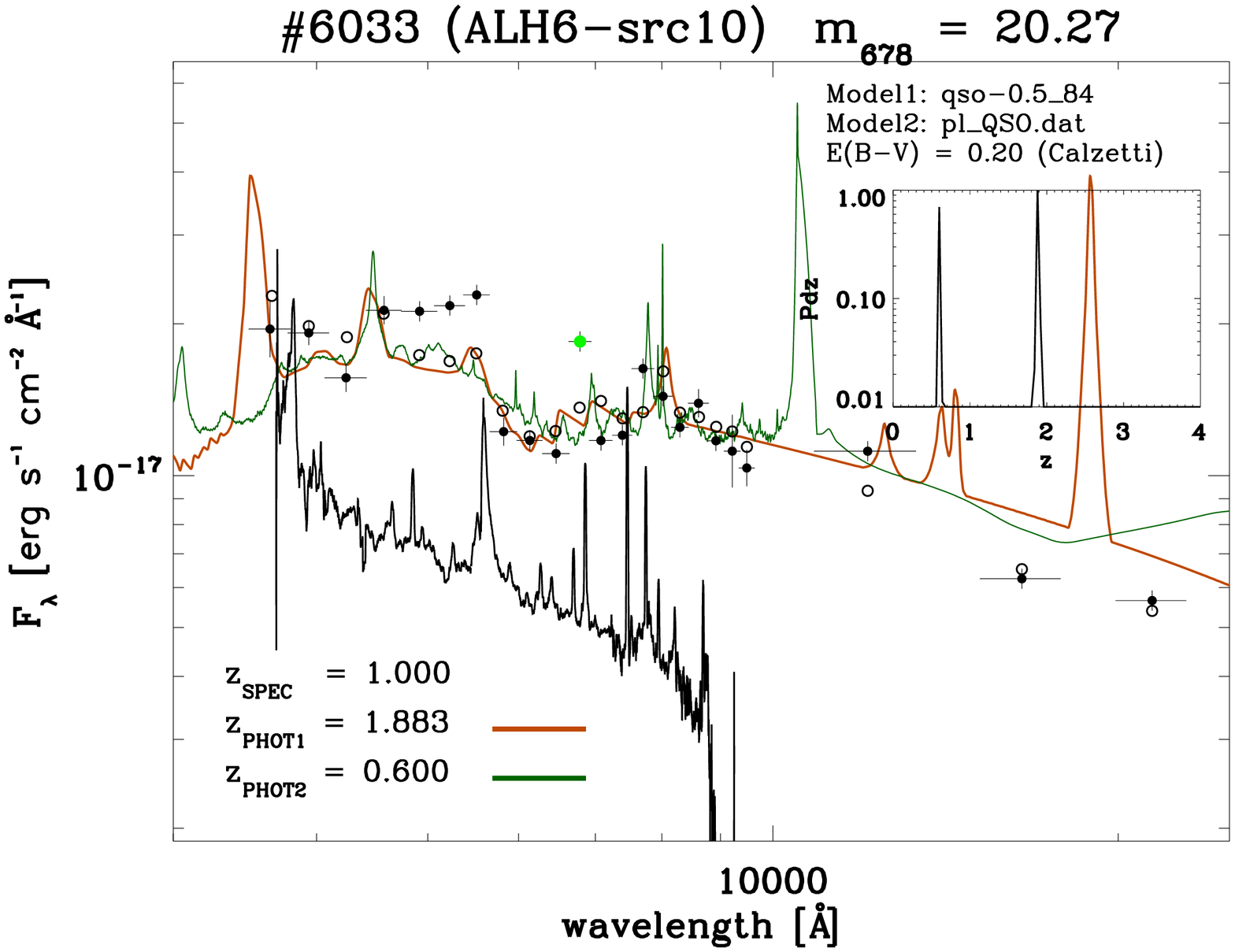}
     \vspace{0.9cm}
      \caption{\scriptsize Example of an object with a strange ALHAMBRA SED
        probably caused by intrinsic variability during the time in which
        photometry was taken. Lines and symbols as in
        Fig.\,\ref{Fig:SEDfit_examples2c}. We also indicate with a green line
        a possible secondary solution with $z_{PHOT} = 0.6$. As a comparison,
        we include the observed spectra with a continuous black solid
        line. Since no flux calibration is provided with the observed spectra,
        we scale the flux accordingly to fit the plot scale.  The observed
        spectra has been smoothed using a 20 pixel box.} 
     
      \label{Fig:variability_src}
   \end{figure}
%


%
\begin{table*}[ht]
\caption{Photo-$z$ accuracy comparison in different cosmological fields}
\label{table:fields-photoz}
\centering
\tiny
\begin{tabular}{c c c c c c c c c c}
\multicolumn{10}{c}{GALAXIES} \\
\hline\hline     
Survey & 
\# Sources\tablefoottext{a} & 
\# Bands\tablefoottext{b} & 
$\lambda$--range & 
Depth & 
Code & 
\# Temp.\tablefoottext{c} &
$\sigma$ &
$\eta$\tablefoottext{d} 
& Ref.\tablefoottext{e} \\ 
\hline       

  COSMOS      & 4148 & 30 (12) & UV--IRAC & $i_{\mathrm{AB}}^* \le22.5$ & LePhare & 31 & 0.007 & $<1$\,(0.15)& (1)\\ 
  GOODS-N/HDF-N &$\sim$130& 7(--)&$UBVIJHK$&	$I_{814}\le25$	&BPZ	&6&0.06	&$\sim$1.0	& (3, 4) \\
  MUSYC\,(all) & 2551 & 32 (18) & Opt--8.0\,$\mu$m& $R\le26$ & EA$z$Y &semi-analytical& $\sim$0.010 &
      $\sim$5.0\,(0.10) & (6)\\ 
  SWIRE (all)     & 5982 & 6--10 (--) & 0.36-4.5\,$\mu$m& $r<23.5$&ImpZ-2& 9 & 0.035 & $\sim$1.0 (0.15) & (7)\\
  COMBO-17   & 813 & 17 (12)& 0.35--0.93\,$\mu$m & $R<24$ & -- &PEGASE lib.& $\sim$0.07 & $\sim$2.0\,(0.15) & (8) \\
  NDWFS--Bo\"otes & 14448 & 13 (--)&UV-Opt-NIR-MIR &$R<26$&$\chi^2$ min.&4 LRT$^g$& 0.041  & 5.0 & (10) \\
  Lockman Hole & 209 & 21 (--) & FUV-Opt-NIR & $R_c<22.5$ & LePhare & 31 &
    0.034 & 10.0 (0.15) & (11) \\

\hline                                   
\\
\\
\multicolumn{10}{c}{AGN} \\
\hline\hline  
  XMM-COSMOS$^f$ & 236  & 30 (12) & UV--IRAC & $i_{\mathrm{AB}}^* \le22.5$ & Le Phare & 30/31& 0.013 & 7.2\,(0.15)& (2)\\
  C-COSMOS$^f$ & 236  & 30 (12) & UV--IRAC & $i_{\mathrm{AB}}^* \le22.5$ & Le Phare & 30 & 0.011 & 5.1\,(0.15)& (2)\\
  MUSYC\,(X--ray) & 236 & 32 (18) & Opt--8.0\,$\mu$m& $R\le26$ & EA$z$Y &1$^g$& 0.012 & 12.0\,(0.12) &(6)\\
  CDF-S       & 446 & 35 (18)  & UV--8.0\,$\mu$m & $z\le26$ & ZEBRA & 265 & 0.01--0.06 & 1.4--8.1\,(0.15) & (5)\\
  SWIRE (QSO)     & 158 & 6--10 (--) & 0.36-4.5\,$\mu$m& $r<23.5$& ImpZ-2 &9& 0.093 & $\sim$33.0 (0.10) & (7)\\
  COMBO-17 (QSO)    & 52 & 17 (12)& 0.35--0.93\,$\mu$m & $R<24$ & -- &1& 0.007 & 17.3\,(0.10) & (8,9) \\
  NDWFS--Bo\"otes & 5347 & 13 (--) &UV-Opt-NIR-MIR &$R<26$&$\chi^2$ min.&4 LRT$^g$& 0.05--0.18 & 5.0 & (10) \\
  Lockman Hole & 90 & 21 (--) & FUV-Opt-NIR & $R_c<22.5$ & LePhare & 30 & 0.069 & 18.9 (0.15) & (11) \\
  \textbf{ALHAMBRA}    & \textbf{170} & \textbf{23 (20)} & \textbf{Opt--NIR} & \textbf{$m_{678}\le23.5$}$^h$ &
      \textbf{LePhare}&\textbf{50}& \textbf{0.009} & \textbf{12.3\,(0.15)}& (12)\\

\hline

\end{tabular}
\tablefoot{\scriptsize \\ \tablefoottext{a}{Number of sources with
    spectroscopic redshift used for photo-$z$ calibration.}
  \\ \tablefoottext{b}{Maximum number of photometric bands used for photo-$z$
    determination. In parenthesis, the number of those filters that have a
    narrow or medium passband.} \\ \tablefoottext{c}{Number of extragalactic
    templates used for photo-$z$ determination.} \\ \tablefoottext{d}{Percent
    fraction of outliers. Outlier threshold criteria defined as $|\Delta
    z|/(1+z_{\mathrm{spec}})$ in parenthesis.}
  \\ \tablefoottext{e}{References: (1)=Ilbert et al. (2009); (2)=Salvato et
    al. (2011); (3)= Ben\'itez et al. (2000); (4) = Coe et al. (2006); (5)=Luo
    et al. (2010); (6)=Cardamone et al. (2010); (7)=Rowan-Robinson et
    al. (2008); (8)=Wolf et al. (2004); (9)=Wolf et al. (2008); (10)= Assef et
    al. (2010); (11)= Fotopoulou et al. (2011); (12)= This work.}
  \\ \tablefoottext{f}{We refer here to the sub-sample of point-like/variable
    sources from the total XMM COSMOS catalog (the \textit{QSOV} sample).}
  \\ \tablefoottext{g}{Low Resolution Templates. The photometric redshift code
    allows for an interpolation between all templates (galaxy + AGN).}
  \\ \tablefoottext{h}{This magnitude corresponds to roughly the broadband
    Sloan filter $r \sim 24$. }
}
\end{table*}
%

4) \textit{Intrinsic variability}. Seven sources ($\sim$33\% of all the
outliers; Fig.\,\ref{Fig:SMC-accuracy2} \#'s 4003, 4012, 4039, 4043, 4054,
6033 and 6042) have unusual ALHAMBRA SEDs that are incompatible with the 
template database, indicating that variability is probably playing an important role. 
None of the sources display signs of either blending or a close companion that might 
contaminate their photometry. Figure\,\ref{Fig:variability_src} shows one of these
sources and illustrates the disagreement in the continuum and emission
features between the ALHAMBRA photometry and the observed spectra. An estimate
of the intensity of the variability can be obtained from the 5 sources within
the COSMOS field using the catalog published by Salvato et al. (2011).  All
5 sources display flux variability of $\Delta m>0.35$ and in some cases as large
as $\Delta m\sim0.80$. The importance of the variability correction for the
photo-$z$ computation of BLAGN and QSOs has been reported in the past by
several authors (Wolf et al. 2004, Salvato et al. 2009). Unfortunately, owing 
to the observing strategy of ALHAMBRA, it is impossible to set the photometry 
to a common epoch.

For only one source (\#4083) does its incorrect photo-$z$ solutions
seem unclear as it has a fairly bright ($m_{678}=18.54$) power-law spectrum
with low variability ($\Delta m<0.2$ from Salvato et al.  2011).

In summary, we found that $\sim$10\% of the outliers can be explained by
their less precise photometry, $\sim$38\% are caused by emission-line
misclassification, $\sim$14\% shown clear evidence of an incorrectly
assigned spectro-$z$, and the remaining $\sim$33\% have significant intrinsic
variability.

\subsection{Performance comparison with other surveys}

Our results are compared to those already published in the
literature in other cosmological fields. Table\,\ref{table:fields-photoz}
summarizes the main characteristics of the photo-$z$ accuracy for
several relevant cosmological surveys, including the wavelength range covered,
number of photometric bands used, number of sources in the spectroscopic
sample used to calibrate the photo-$z$, the depth of the photometry, the
photometric code used, the number of templates, the final precision in
terms of the accuracy ($\sigma$), and the fraction of outliers ($\eta$). The
table has been divided into one section for the galaxy and one for the AGN
results. This highlights the accuracy differences between the two populations,
as well as the significantly larger (by factors of $\sim 3-10$) spectroscopic
samples available to the former because AGN represent only $<10$\% of
the extragalactic population.  We note here that the table highlights simply
the differences in the wavelength coverage and filter type used by
each survey. The reader must be aware of the caveats in this comparison,
namely the different depths, source populations, methodology, spectroscopic
sample used, and the computation of the accuracy estimates.

The benefit of a continuous optical coverage with medium--band photometry is
clearly illustrated in Tab.\,\ref{table:fields-photoz}, where we compare broadband
surveys such as SWIRE and HDF-N with medium--band surveys like MUSYC, COMBO-17,
and ALHAMBRA itself.  The precision of the results are particularly higher for the BLAGN
population, where the accuracy increases by a factor of $\sim$10 and the
fraction of outliers decreases by a factor of 3 or more.

A more interesting comparison is that between ALHAMBRA and the surveys
that include a certain number of medium optical filters and different
wavelength coverages. Since the ALHAMBRA survey is a natural successor 
to the COMBO-17 survey (Wolf et
al. 2003), it is mandatory to compare the performance of its
photometric system with that of COMBO-17 in the E-CDFS field.  We found that, for
BLAGN/QSO, ALHAMBRA is only marginally worse than the latest recalibration of
the COMBO-17 data (Wolf et al. 2008) in terms of accuracy (0.009 vs. 0.007). On
the other hand, it has a smaller fraction of outliers (12.3\% vs. 17.3\%). The
narrower medium--band passbands of COMBO-17 (FWHM of $\sim$200\,\AA \, vs.
$\sim$300\,\AA) may be held responsible for the slight accuracy advantage, but
also introduce a highly redshift-dependent selection function owing to the 
non-continuous optical coverage of the filters, as in the case of ALHAMBRA. 
This continuous, non-overlapping coverage of the optical by the
ALHAMBRA filters, together with the addition of NIR broadband photometry,
has helped to reduce the number of catastrophic failures.  Furthermore, our
results are obtained without applying the variability correction made to
COMBO-17 data.

As part of the Multiwavelength Survey by Yale--Chile (MUSYC; Gawiser et
al. 2006), photo-$z$ estimates results were published by Cardamone et
al. (2010) in the E-CDFS using the deeper, more extensive, broad-band
photometry and also including photometry from 18 medium-band filters taken
with the Subaru telescope. The performance comparison between the ALHAMBRA and
MUSYC X--ray (AGN-dominated) population yields identical values. This result
is unsurprising given that the definition of the MUSYC medium--band
photometric system is very similar to that of ALHAMBRA, i.e. continuous,
non-overlapping medium--band filters. Nevertheless, we note that, given the
nature of the E-CDFS observations, the results obtained by MUSYC extend to much fainter
optical magnitudes ($R\sim26$).  Photometric redshifts estimates are also
available in this field for the sources detected by the \textit{Chandra} 2\,Ms
exposure in the 436\,arcmin$^2$ CDF-S (Luo et al. 2010).  Although the highest
accuracy and outlier fraction found by the authors is as low as 0.01 and 1.4\%
respectively, they also found that the strong dependence of this result on
the spectroscopic sample, used to correct their SED library, could increase the
dispersion and outlier fraction of the result to $\sigma_{NMAD}=0.059$ and
$\eta=8.1$\%. A larger dispersion would be expected as their analysis also
extends to a much fainter population ($z<26$) as a result of the depth of the
\textit{Chandra} observations ($\sim2\times
10^{-17}$\,erg\,cm$^{-2}$\,s$^{-1}$ in the 0.5--2.0\,keV band).  The Luo et
al. results have been included under the AGN section of
Table\,\ref{table:fields-photoz}, although given the depth of the
\textit{CHANDRA} observations, an important fraction of sources is expected to
be dominated by intrinsically faint AGN, starbursts, or even normal galaxies.

  \begin{figure*}[!t]
  \centering
    \includegraphics[width=6.4cm,angle=90,clip=true,trim=0 10 0 15]{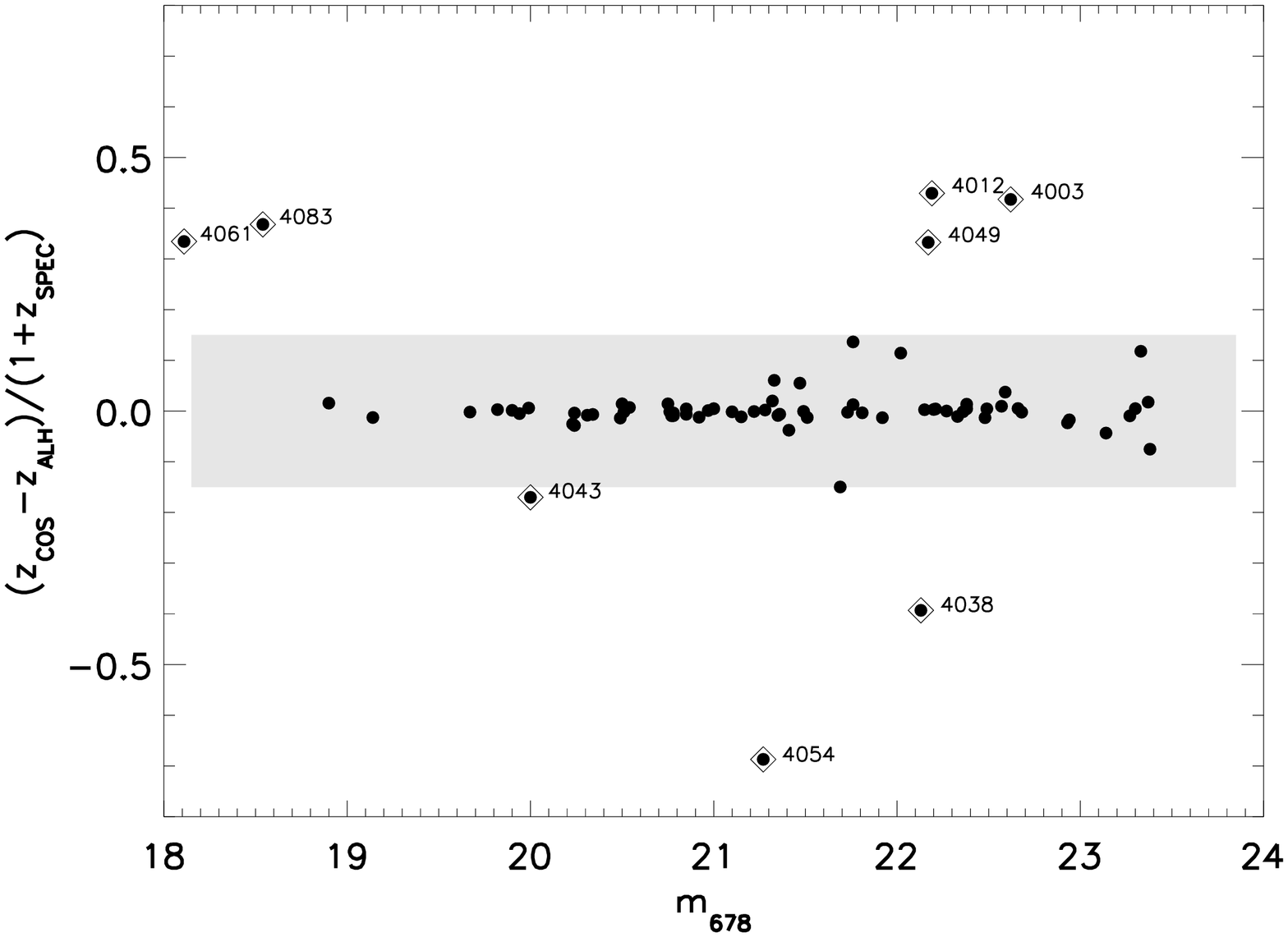}
    \includegraphics[width=6.4cm,angle=90,clip=true,trim=0 10 0 15]{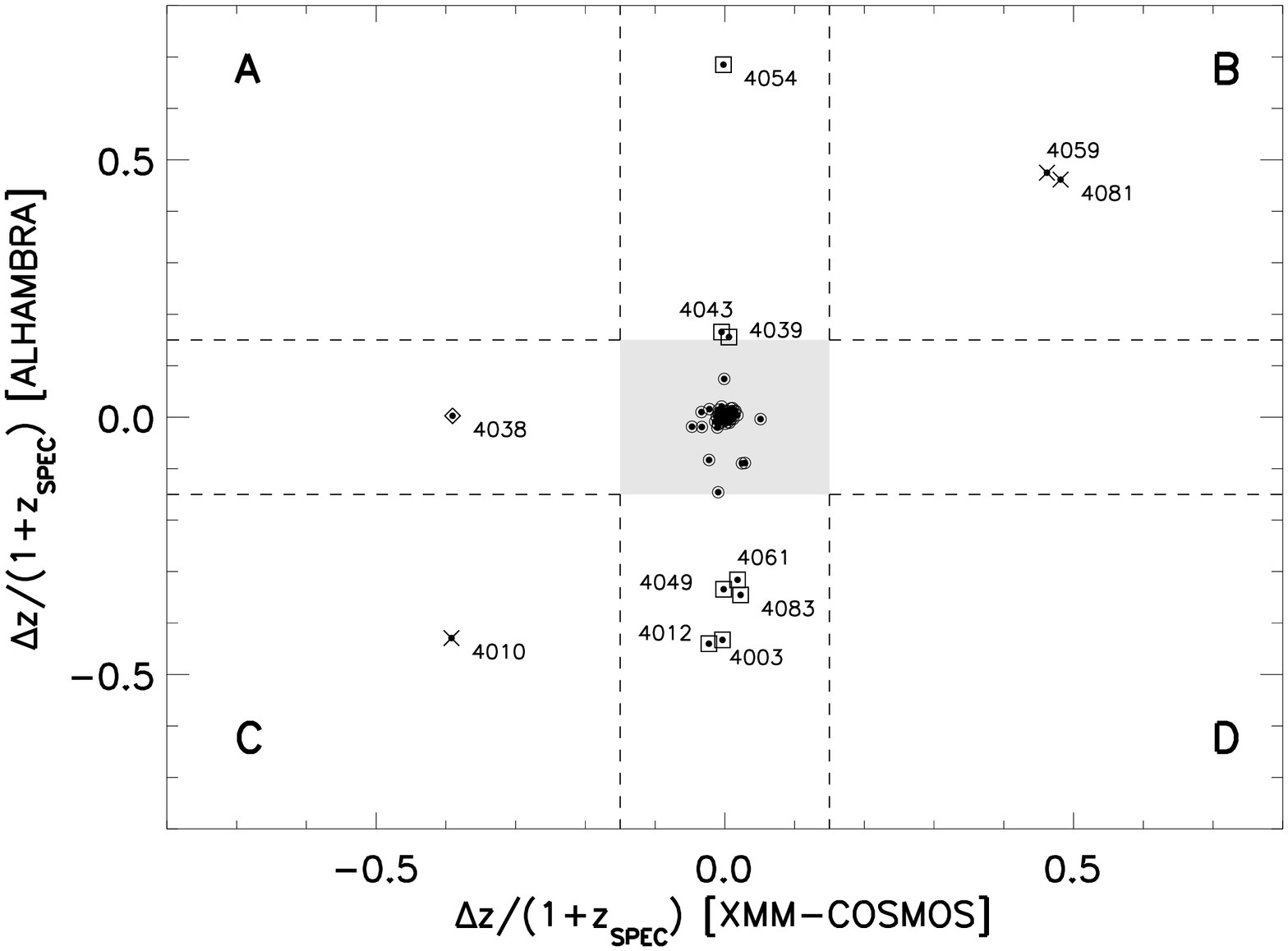}
      \caption{\scriptsize Comparison of the photo-$z$ solutions for the 77
        common sources in the XMM--COSMOS field.  \textit{Left}) Differences in
        the photo-$z$ solutions found by the ALHAMBRA and COSMOS photometry as
        a function of the visual magnitudes in the ALHAMBRA filter A678M. Both
        surveys agree for $\sim90\%$ of the sources when
        $(z_{\mathrm{PHOT-COS}} - z_{\mathrm{PHOT-ALH}})/1+z_{\mathrm{SPEC}}
        \le 0.15$, where $z_{\mathrm{PHOT-COS}}$ and $z_{\mathrm{PHOT-ALH}}$
        are the photo-$z$ solutions found in the COSMOS and ALHAMBRA surveys
        respectively.  \textit{Right}) Comparison of the source-by-source
        photo-$z$ accuracy for both surveys. The threshold limit for outliers,
        $|\Delta z|/(1+z_{\mathrm{spec}})\le0.15$, is indicated by vertical
        and horizontal dashed lines for the COSMOS and ALHAMBRA surveys,
        respectively. The central grey square represent the region where the
        two surveys agree, while sources in any of the exterior 4 quadrants (A,
        B, C, and D) have wrong photo-$z$ solution in both COSMOS and ALHAMBRA
        (crosses). Sources within the vertical or horizontal stripes, defined
        by the dashed lines, and not within the central square have either
        $i)$ good--COSMOS and poor--ALHAMBRA solutions (squares) or $ii)$
        good--ALHAMBRA and poor--COSMOS solutions (diamonds).}
     \label{Fig:zphot_comparison_ALH4}
   \end{figure*}
%

The COSMOS survey (Scoville et al. 2007) provides the largest number of
BLAGN/QSOs of our ALHAMBRA sample thanks to its extensive spectroscopic
follow-up (Lilly et al. 2007; Brusa et al. 2010 and references therein). This
spectroscopic follow-up, coupled with the vast multiwavelength information
available in the field, has allowed the computation of precise photometric
redshift for galaxies and AGN (Ilbert et al. 2009; Salvato et al. 2011). Our
photo-$z$ determination and that of COSMOS share a similar methodology, the use of
the same photometric code (\textit{LePhare}) as well as several templates in the
dataset. A direct comparison is therefore possible between their findings and
our results for the BLAGN/QSO population (the \textit{QSOV} sample in Salvato
et al. 2011), allowing us to further quantify the ALHAMBRA results through:
$i$) the comparison of the photometric system of both surveys, i.e. a COSMOS
30 filter--set with a significant number of broad-band filters, including
NUV/GALEX and mid-IR/IRAC photometry but also 12 medium-band filters and 2
narrow-band filters, against the 23 ALHAMBRA filter--set dominated by optical
medium passbands but with no info in the NUV or mid-IR; $ii$) the impact of
the variability correction applied to the COSMOS data.

We found in general good agreement between our photo-$z$ and the fraction of
BLAGN/QSOs published by Salvato et al. (2011) located within the ALHAMBRA
field (77 sources in $\sim$12.5\% of the COSMOS area). The results of this
comparison are presented in Fig.\,\ref{Fig:zphot_comparison_ALH4}. The left
panel shows the overall agreement between both surveys, with 90\% of the
sources (69/77) lying within $(z_{\mathrm{PHOT-COS}} -
z_{\mathrm{PHOT-ALH}})/1+z_{\mathrm{SPEC}} \le 0.15$ over the entire magnitude
range of the sample.  However, this comparison does not establish whether or
when any of the two photometric estimates provide an accurate photo-$z$.  This
information is given in the right panel of Fig.\,\ref{Fig:zphot_comparison_ALH4}, 
where we compare the accuracy found by
COSMOS and ALHAMBRA source by source.  The threshold limits for outliers
($\sigma_{NMAD}\le0.15$) are represented by vertical and horizontal dashed lines
for the COSMOS and ALHAMBRA, respectively. The central grey square contains the
sources for which the two surveys photo-$z$ estimates agree with the measured
spectro-$z$ ($\sim$84\%; 65/77).  For 8 of the 12 sources in disagreement, COSMOS
provides a more accurate solution (\textit{squares} in
Fig.\,\ref{Fig:zphot_comparison_ALH4}), while 1 source has a more reliable estimate
in ALHAMBRA (\textit{diamond}). Of the 8 sources with more accurate COSMOS
solutions, 5 need a strong variability correction according to Salvato et al. (2009; see
Sect.\ref{subsection: outlier nature}), which may explain our wrong
estimates, 2 sources have the line misclassification uncertainty described
in Sect.\,\ref{subsection: outlier nature} (\#4061 and \#4083) and another
one has large photometric uncertainties in ALHAMBRA (\#4049). Finally, in both
surveys, 3 sources have photo-$z$ estimates that inconsistent with their spectro-$z$
(quadrants B and C in Fig.\,\ref{Fig:zphot_comparison_ALH4}). We note that these 
photo-$z$ estimates are very similar for 3 of the sources
(\# 4059, 4081 and 4010) and that the redshifts of these sources are based on a
single line spectrum. For two of them, there seems to be an incorrect
measurement of the spectro-$z$ (see Sect.\,4.3).  For the third source, the
photo-$z$ solution is degenerate, showing two peaks of similar intensity in
the Pd$z$ distribution, and the method selects the incorrect photo-$z$.

To summarize, we observe that the accuracy of the ALHAMBRA photometric
redshift determination for the BLAGN/QSO population is, as expected, much
better than any broad--band survey and similar to any recent photometric
survey including medium passband filters. This result is obtained without
the advantage of the larger wavelength coverages of other surveys and thanks to the ALHAMBRA
filter selection in the optical and NIR. The fraction of outliers is
comparable to or better than other cosmological surveys that include no
variability correction. The importance of the variability correction for
BLAGN/QSOs (Wolf et al. 2004; Salvato et al. 2009) is highlighted by our
larger outlier fraction (by a factor of $\sim$2) compared to a survey such as
COSMOS, whose observation strategy allowed all its photometry to be set to a
common epoch.

\subsection{The impact of the NIR photometry}

The need for NIR photometry in order to increase the accuracy of photo-$z$ and
reduce the number of catastrophic failures for normal and starburst galaxies
has been demonstrated by several authors (Connolly et al. 1997;
Fern\'andez-Soto et al. 1999; Rowan-Robinson et al. 2003; Ilbert et
al. 2006). The subsequent improvement (with an increase in accuracy and a reduction in
the outlier fraction by factors of $\sim$2-3; Ilbert et al. 2006) is caused by
these photometric bands providing a tighter constraint of photo-$z$
solutions as they sample the peak emission of the old stellar population
($\sim 1\,\mu$m rest-frame) in the $z\sim$[0,1] redshift range. At redshifts
above $z\sim1.5-2$, the peak stellar emission moves out the wavelength range
covered by the \textit{JHK$_S$} filters and longer wavelength photometry (e.g.
\textit{Spitzer}/IRAC) is required to sample this emission. In the case of the
AGN population, Salvato et al.  (2009, 2011) illustrated that, when only
broad-band imaging is available, sufficiently high photo-$z$ precision and an outlier
fraction suitable to scientific analysis can only be obtained with the
addition of NIR photometry.

We investigated here the importance of NIR photometry to the photo-$z$
solutions of BLAGN and QSOs when medium-band photometry is available in the
optical. To do so, we proceeded as described in Sect.\,\ref{section:QSO-photoZ-method} 
without using the $J$, $H$, and $K_S$ broadband filters. 
The results of the analysis are summarized in
Table\,\ref{table:Fit_results} (columns \#5 to \#8) for the SEL and MEL
solutions. We found that the impact of the NIR photometry is negligible in
terms of accuracy (with a marginal decrease of 0.2--0.3\%) but does have 
influence on the fraction of outliers, which increases by 3-5\% depending on
the set of extinction laws considered during the fit. The larger outlier
fraction is dominated by sources with degenerate solutions (showing two or
more peaks in the Pd$z$ distribution) that can be constrained by the NIR
photometry. On the other hand, the distribution of the SEDs and the required
extinction is also very similar to the solution with and without $JHK_S$
photometry. This result is important to unveiling and characterizing the BLAGN
population of future very large ALHAMBRA-like optical photometric surveys such 
as the JPAS\footnote{http://j-pas.org/} survey or to planning medium-band optical
photometric follow-ups to characterize the population of the all-sky X--ray
survey to be carried out by the future
eROSITA\footnote{http://www.mpe.mpg.de/erosita/} mission (Cappellutti et
al. 2011), as detailed below.

We conclude that, although the use of NIR photometry helps us to alleviate the
fraction of outliers, its importance for BLAGN/QSO photo-$z$ determination (in
the $z=[0,4]$ redshift range) is significantly smaller than for the normal and
starburst population when a continuous coverage of the optical with
medium-band photometry is available. Nevertheless, the importance of $JHK_S$ 
photometry obviously increases with redshift and in particular at $z>5$, where the
limited number of optical detections due to the continuum depression bluewards of
Ly$\alpha$ make the NIR bands necessary to constrain the continuum slope
of BLAGN/QSOs and their photo-$z$ solutions.

\subsubsection{A practical case}

In the future, several surveys will provide large samples of BLAGN and
QSOs. One of these surveys is to be carried out by the eROSITA mission. eROSITA
will map all the sky more than one order of magnitude deeper in X--ray flux
($F[2-10\,\mathrm{keV}]\sim10^{-13}$\,erg\,cm$^{-2}$\,s$^{-1}$) than the
previous \textit{ROSAT} All Sky Survey. The deep part of the survey will cover
$\sim$200\,deg$^2$ down to a flux of
$F[2-10\,\mathrm{keV}]\sim4\times10^{-14}$\,erg\,cm$^{-2}$\,s$^{-1}$. In total,
the eROSITA mission is expected to discover more than 10$^6$ new AGN including
large fractions of BLAGN and QSOs given the survey flux limits. To aid the
follow-up of the AGN population uncovered by eROSITA and, in order to obtain
reasonable photo-$z$ solutions ($\sigma_{NMAD}\sim 0.08$ and $\eta\sim22\%$),
Salvato et al. (2011) discussed the need for at least \textit{JHK}
photometry when only 4--5 broad bands are available in the optical, such as those
provided by the very wide-field surveys LSST\footnote{http://www.lsst.org/} and
Pan--STARSS\footnote{http://pan-starss.ifa.hawaii.edu}.

\begin{figure}[t]
   \centering
    \includegraphics[width=6.6cm,angle=90]{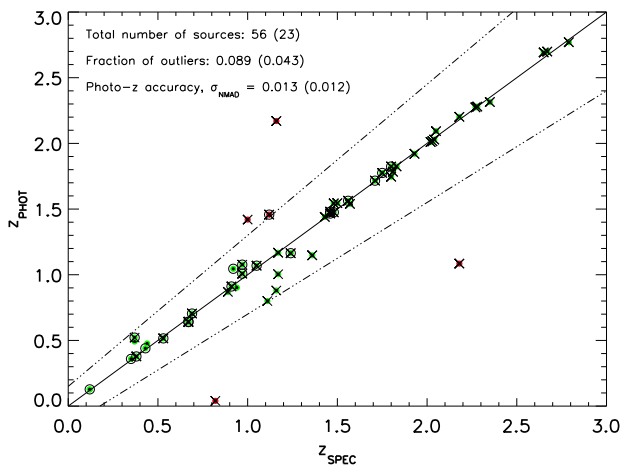}
      \caption{\scriptsize Photometric and spectroscopic redshift comparison
        for the ALHAMBRA/XMM-COSMOS sources with eROSITA deep survey fluxes
        ($F[0.5-2\,\mathrm{keV}] >
        4\times10^{-15}$\,erg\,cm$^{-2}$\,s$^{-1}$).  Symbols color-coded as
        in Fig.\ref{Fig:SMC-accuracy2}.  Open circles indicate the sources
        with a X--ray flux above $F[0.5-2\,\mathrm{keV}] >
        4\times10^{-15}$\,erg\,cm$^{-2}$\,s$^{-1}$ (eROSITA all sky). Crosses
        highlight sources with spectroscopic BLAGN classification, while
        open-squares indicate which sources are classified as such with the
        ALHAMBRA photometry.  }
     \label{erosita_zphot}
   \end{figure}

Assuming an ALHAMBRA-like survey covering the entire sky, we can estimate the
number of BLAGN/QSOs that could be photometrically identified in eROSITA
following the methodology described here. For this purpose, the XMM--COSMOS
field serves as a benchmark of the accuracy achievable with such an approach by cutting the
XMM-COSMOS X--ray catalog (Brusa et al. 2010) to the fluxes to be reached by
the eROSITA all-sky and deep surveys as well as the final magnitude cutoff of
the ALHAMBRA photometry ($r\sim25$). This X--ray selection yields a total of
65 (56 with spectro-$z$) sources within the ALHAMBRA-COSMOS region for the
deep selection and 24 (23 with spectro-$z$) for the shallow all--sky
selection. Photo-$z$ solutions were computed for these sources using the
ALHAMBRA photometry and the results are presented in
Fig.\,\ref{erosita_zphot}. The photo-$z$ accuracies obtained
($\sigma_{\mathrm{NMAD}}=0.013$, $\eta=8.9\%$ for eROSITA-deep;
$\sigma_{\mathrm{NMAD}}=0.012$, $\eta=4.3\%$ for eROSITA-shallow) are
significantly higher than those for broad-band photometry and similar to the
results discussed in Sect.\,\ref{section:Results}. The spectroscopic sample is
dominated by BLAGN/QSOs (82\%, 46 out of 56) for which the ALHAMBRA photometry
is able to recover a correct BLAGN/QSO classification (templates from \#29 to
\#49 in Fig.\,\ref{templates}) for $78\%$ (36/46) and $90\%$ (17/19) of the
deep and shallow samples, respectively. The extrapolation of this efficiency from the
ALH-4/COSMOS area (0.25\,deg$^2$) to the eROSITA deep (200\,deg$^2$) and all-sky
extragalactic (20000\,deg$^2$) surveys show that an ALHAMBRA-like photometric
survey can select and provide high accuracy photo-$z$ solutions for more than
$4\times10^4$ and $1\times10^6$ BLAGNs, respectively.

Unfortunately, the telescope time that would be required to carry out a 
survey such ALHAMBRA, even in the smaller deep regions of eROSITA, is too long 
for any publicly available observatory, i.e. the time to execute an ALHAMBRA-like survey in
200\,deg$^2$ will be $\sim$10 times longer than that of ALHAMBRA itself, roughly 40
years.  On the other hand, a dedicated telescope facility optimized for large-area
photometric surveys would do the job in a fraction of the time. This is
the case of the planned Javalambre PAU (Physics of the Accelerating Universe)
Astrophysical Survey (JPAS, Ben\'itez et al. 2009). This photometric survey,
will be carried out at the Javalambre observatory with a 2.5m telescope and a
5\,deg$^2$ FOV camera, and will sample 8000\,deg$^2$ of the northern sky over four
years, starting in late 2013.  The JPAS survey will take the photo-$z$
precision to the next level with an expected accuracy of $\sigma=0.003$ thanks 
to its 54 narrow-band optical filters (FWHM$\sim$100\AA). 
The classification and photo-$z$ estimates of BLAGN and QSOs
will be significantly higher in the common region covered by JPAS and eROSITA
with respect to ALHAMBRA. This is due to the narrower JPAS filter set and its
ability to correct for variability. On the basis of the above ALHAMBRA/eROSITA
results, we expect JPAS to select and provide very accurate photo-$z$ for at
least $4\times10^5$ BLAGN/QSOs in the JPAS/eROSITA common area.

\section{Conclusions and future work}

We have explored the ability of the ALHAMBRA survey
photometry to assign very accurate photometric redshifts to a population of
BLAGNs and QSOs. We have achieved a precision better than 1\% using a catalog of
170 spectroscopically identified BLAGN/QSO in the ALHAMBRA fields. This
precision is similar to the previously published photo-$z$ accuracy for BLAGN/QSOs
in surveys that make use of medium-band optical photometry. This has been
possible despite the limited wavelength coverage of ALHAMBRA relative to other 
surveys and thanks to its photometric system definition (consisting in 20 
continuous, non-overlapping optical medium-band filters plus the 3 NIR 
broadbands $J$, $H$, and $K_s$).

We have used the publicly available code \textit{LePhare} to derive our photo-$z$ 
solutions by means of fitting our photometry to a template database of 
normal and starburst galaxies, type-1/2 Seyfert, QSOs, and stars. Our 
treatment has included a correction for systematic offsets between the 
different photometric bands and the template database, Galactic
extinction, absorption by IGM, and the possibility of intrinsic
reddening by adopting various extinction laws. In addition to the excellent
accuracy of out photo-$z$, our analysis has demonstrated that:

\begin{enumerate}
 \item The medium-band filter set used by the ALHAMBRA photometry is able to
   detect the emission and absorption features that characterize QSO optical
   spectra in the redshift interval $0< z \le 4$.
 \item We have been able to easily differentiate QSO emission from stellar 
    emission over the entire QSO redshift range from 0 to 3.
 \item In the redshift interval $0< z \le 4$ and $m_{678} \leq 23.5$, we have
   characterized the nature of 87.7\% of the sources providing a correct SED
   type and robust photo-$z$. The fraction of outliers (12.3\%) shows a clear
   correlation with magnitude with $\sim$70\% of the outliers being located at
   $m_{678} > 21.0$.
 \item The most probable reasons for our fraction of outliers
   are: $i$) the faintness or poor photometric quality of some
   sources, $ii$) line misclassification, $iii$) the intrinsic variability of
   the AGN population (for which the ALHAMBRA photometry applied no
   correction), and $iv$) the possibility of an incorrect spectro-$z$ assignment
   owing to the limited wavelength coverage of the spectra of some objects.
 \item Near-IR photometry is not fundamental to constrain the photo-$z$
   solutions of BLAGN and QSOs, at least in the redshift interval $0<z<4$, if
   the optical regime is covered by a continuous medium-band filter set such as
   that of ALHAMBRA. This result is relevant to the design of future optical
   follow-ups of surveys with a large fraction of BLAGN, as in the case of either
   X--ray or radio surveys.

\end{enumerate}

We therefore conclude that the analysis performed here validates the
feasibility and accuracy of the photo-$z$ determination for a large number of
sources with the ALHAMBRA limited (but well-defined) set of filters. In the
particular case of BLAGN and QSOs, our methodology and results suggest the
potential ability of the ALHAMBRA photometry to detect these sources and build
a large and unbiased BLAGN/QSO database.  The publication of such a catalog
as well as the precise estimate of its inherent incompleteness necessary to
derive meaningful statistical properties (e.g. number counts and luminosity
functions) are left to a forthcoming paper.

\begin{acknowledgements}
      The authors will like to thank Mara Salvato for her helpful advice and for providing
      templates, solution details, and variability corrections for the
      photo-$z$ determination of \textit{XMM-Newton} sources in the COSMOS
      field. This work has made use of the TOPCAT tool (Taylor 2005). We thank
      C\'ecile Cartozo for the careful reading of this manuscript and an
      anonymous referee for constructive comments and helpful suggestions that
      improved the quality of this paper. Part of this work was supported by
      Junta de Andaluc\'ia, through grant TIC-114 and the Excellence Project
      P08-TIC-3531, and by the Spanish Ministry for Science and Innovation
      through grants AYA2006-1456, AYA2010-15169, AYA2010-22111-C03-02, and 
      AYA2011-29517-C03-01.     
\end{acknowledgements}

\end{document}